\documentclass[journal]{IEEEtran}
\usepackage{amsmath}
\usepackage{mathtools, nccmath}
\usepackage{amsfonts}
\usepackage{textcomp}
\usepackage{graphicx}
\usepackage{gensymb}
\usepackage[font={footnotesize}]{caption}
\usepackage{subfig}
\usepackage{lipsum}
\usepackage{etoolbox}
\usepackage{xcolor}
\usepackage{array}
\usepackage{tabulary}
\usepackage{booktabs}
\usepackage{multirow}
\usepackage{subfig}
\usepackage[hang, flushmargin]{footmisc}
\usepackage{longtable, tabularx, ltablex}
\usepackage{float}
\usepackage{threeparttable}
\usepackage{cite}
\renewcommand{\thetable}{\arabic{table}}
\def\tablename{Table}
\def\BibTeX{{\rm B\kern-.05em{\sc i\kern-.025em b}\kern-.08em
    T\kern-.1667em\lower.7ex\hbox{E}\kern-.125emX}}
\begin{document}
\title{Coded Aperture and Compton Imaging for the Development of $^{225}$Ac-based Radiopharmaceuticals}
\author{Emily Frame, Kondapa Bobba, Donald Gunter, Lucian Mihailescu, Anil Bidkar, Robert Flavell, and Kai Vetter
\thanks{Manuscript received XX-XX-XX. This work was supported by the Department of Energy National Nuclear Security Administration through the Nuclear Science and Security Consortium under Award Number DE-NA0003180.}
\thanks{E. Frame and K. Vetter are with the Department of Nuclear Engineering, University of California Berkeley, Berkeley, CA~94502 (email: eaframe@berkeley.edu).}
\thanks{K. Bobba, A. Bidkar, and R. Flavell are with the Department of Radiology and Biomedical Imaging, University of California San Francisco, San Francisco, CA~94143.}
\thanks{D. Gunter is with Gunter Physics, Inc., Lisle, IL~60532.}
\thanks{L. Mihailescu is with Ziteo Medical, Inc., Pleasant Hill, CA~94523.}}

\maketitle

\begin{abstract}
\label{sec:abs}
Targeted alpha-particle therapy (TAT) has great promise as a cancer treatment. Arguably the most promising TAT radionuclide that has been proposed is $^{225}$Ac. The development of $^{225}$Ac-based radiopharmaceuticals has been hampered due to the lack of effective means to study the daughter redistribution of these agents in small animals at the preclinical stage. The ability to directly image the daughters, namely $^{221}$Fr and $^{213}$Bi, via their gamma-ray emissions would be a boon for preclinical studies. That said, conventional medical imaging modalities, including single photon emission computed tomography (SPECT) based on pinhole collimation, cannot be employed due to sensitivity limitations. As an alternative, we propose the use of both coded aperture and Compton imaging with the former modality suited to the 218-keV gamma-ray emission of $^{221}$Fr and the latter suited to the 440-keV gamma-ray emission of $^{213}$Bi. This work includes coded aperture images of $^{221}$Fr and Compton images of $^{213}$Bi in tumor-bearing mice injected with $^{225}$Ac-based radiopharmaceuticals. These results are the first demonstration of visualizing and quantifying the $^{225}$Ac daughters in small animals via coded aperture and Compton imaging and serve as a stepping stone for future radiopharmaceutical studies.
\end{abstract}


\section{Introduction}
\label{sec:intro}
\IEEEPARstart{T}{argeted} delivery of alpha-particle-emitting radioisotopes has great potential as a cancer therapy. The short range and high linear energy transfer (LET) of alpha particles enables highly-selective and effective killing of tumors while sparing normal tissues \cite{pouget}. In theory, the efficacy of the delivered dose can be enhanced further by employing radioisotopes with decay progeny that also emit alpha particles. One of the more attractive so-called nanogenerators that has been proposed is $^{225}$Ac \cite{mcdevitt, pros1}. Actinium-$225$ is a relatively long-lived radiometal with a half-life $t_{1/2}$ of $10$~days and decays via a sequence of six short-lived daughters to stable $^{209}$Bi. The predominant decay pathway of $^{225}$Ac yields four alpha particles with contributions from the daughters $^{221}$Fr ($t_{1/2}=4.90$~min), $^{213}$Bi ($t_{1/2}=45.6$~min), and $^{213}$Po ($t_{1/2}=4.2$~$\mu$s).

Despite great promise, the development of $^{225}$Ac-based therapies has been hampered, because there has been no effective means to study the daughter redistribution \cite{de2015}. Alpha-particle-emitting daughters, once formed, can possibly break free from the chelator of the radiopharmaceutical due to a high recoil energy and different chemical properties. If the daughters are generated and retained inside the cancerous cells after internationalization, they can contribute to the cytotoxic effect. Otherwise, free daughters, produced either on the surface of the target cell or during circulation of the radiopharmaceutical, can travel to healthy organs; thereby resulting in unwanted toxicity. In the case of $^{225}$Ac, the redistribution of its longer-lived daughters, namely $^{221}$Fr and $^{213}$Bi, evokes the most concern.

A typical biodistribution study starts with the administration of the drug under investigation in multiple mice. These mice are subsequently sacrificed at different time points post-injection. Following each sacrifice, organs are dissected, weighed, and counted (oftentimes using a scintillation counter) to assess the drug accumulation at the given time interval. This entire approach is time-consuming, laborious, and arguably unethical as a large number of animals are required. Furthermore, postmortem analysis prohibits the ability to monitor the biodistribution in individual animals, and organ counting is susceptible to selection bias because unforeseen drug accumulation in unharvested organs is missed.

Overcoming the limitations of invasive biodistribution studies is the primary motivation of small-animal molecular imaging. The non-invasive nature of imaging offers the exciting possibility of evaluating the time-dependent behavior of radiopharmaceuticals \textit{in vivo}. This enables the collection of a complete set of biodistribution data from a single animal, thereby reducing biological variability and improving data quality. In nuclear medicine, the molecular imaging modalities that have received the greatest attention are positron emission tomography (PET) and SPECT. While PET and SPECT have been employed to great effect in the past, there are several considerations that confound their employment in the preclinical evaluation of the $^{225}$Ac daughters, which has been the Achilles heel in developing $^{225}$Ac-based radiopharmaceuticals.

PET scanners require positron emission, which does not appear in the $^{225}$Ac decay scheme. That said, PET can be used to directly image $^{225}$Ac agents if the radiopharmaceutical is modified to accommodate a positron-emitting isotope. Consequently, a variety of positron emitters have been investigated as chemical surrogates for $^{225}$Ac, including $^{68}$Ga, $^{89}$Zr, and $^{134}$Ce. Cerium-134 is one of the more promising surrogates due to its similar chemical properties and half-life of $75.9$~hours, which enables the radiopharmaceutical to be tracked over several days \cite{bailey}. While PET surrogates can provide valuable insight on the biodistribution of $^{225}$Ac agents, they do not provide information about the daughter redistribution.

In principle, SPECT scanners can directly image radionuclides if photon emissions accompany their decay. Indeed, the $^{225}$Ac decay scheme includes gamma-ray emissions from the daughters $^{221}$Fr and $^{213}$Bi. The former daughter emits a $218$-keV gamma ray with a branching ratio of $11.6\%$, and the latter emits a $440$-keV gamma ray with a branching ratio of $26.1\%$.

SPECT traditionally employs either a pinhole or parallel-hole collimator, both of which are known to achieve high spatial resolution on a sub-millimeter scale \cite{mcelroy, van}. However, collimator-based imagers have a limited photon energy range at which they are effectively operational. At energies above about $300$~keV, these systems experience a degradation of response due to the increased likelihood of unwanted photon transmission through the collimator. This eliminates the potential of effectively utilizing the $440$-keV gamma-ray emission of $^{213}$Bi. Furthermore, even at energies below $300$~keV, conventional SPECT systems have poor imaging sensitivity due to a collimator-driven trade-off between sensitivity and resolution. The sensitivities of dedicated high-sensitivity multi-pinhole collimators rarely exceed values on the order of $0.1\%$ for $^{99m}$Tc \cite{van, deleye}. The tradeoff between sensitivity and resolution is not ideal for preclinical studies as these studies involve imaging low doses of radiation on a small scale. In the case of alpha-particle-based radiopharmaceuticals, studies would require particularly low amounts of activity to be injected due to the high efficacy of alpha particles; on the order of $20$~kBq in small mice with only a fraction of that reaching the tumor site \cite{moroz}.

To overcome the limitations of existing small-animal molecular imagers, this work proposes coded aperture and Compton imaging as complementary modalities with the former modality suited to the $218$-keV gamma-ray emission of $^{221}$Fr and the latter suited to the $440$-keV gamma-ray emission of $^{213}$Bi. By incorporating the proposed imaging modalities with a PET scanner, which can provide an estimate of the $^{225}$Ac biodistribution via positron-emitting surrogates, a full picture of the daughter redistribution can be painted.

For energies below a few hundred keV, coded aperture imagers are advantageous. First introduced by Ables \cite{ables} and Dicke \cite{dicke}, coded apertures decouple the dependence of resolution on sensitivity, providing the maximum possible sensitivity among collimator-based imagers. The main idea behind the coded aperture design is to increase photon acceptance by opening many small pinholes as opposed to widening a single one. The pinholes are optimally arranged to form the coded aperture or mask. The mask situated in front of a position-sensitive detector collectively constitutes the coded aperture imager. A source within the field-of-view of the imager projects a distinct pattern, unique to its distribution, on the detector plane. The detector measurement is then decoded to determine the gamma-ray distribution.

For energies above a few hundred keV, Compton-scattering-based instruments are appealing as they rely on the dominant interaction process, namely Compton scattering, at these energies, and they do not require a collimator that would otherwise decrease the instrument sensitivity. In general, Compton imaging requires an incident photon to interact at least twice within a single or multiple detectors. By measuring the energy depositions and positions of the first two interactions, a so-called Compton cone can be back-projected into the image space using kinematic relations. The surface of this cone contains all possible positions from which the incident photon could have originated. By acquiring many Compton events and analyzing the intersection of the associated cones, the gamma-ray distribution can be determined \cite{todd, phillips}.

There are several critical design considerations in applying both the coded aperture and Compton imaging concepts to small-animal applications. Because these applications involve weak and small-scale radiation distributions, they require an imager that exhibits both high sensitivity and resolution. To maximize both of these parameters, the distance between the small animal and imager should be minimized. Under such near-field conditions, coded apertures are subject to more severe collimation and magnification effects due to the diverging nature of the incident gamma rays \cite{accorsi}. Design features such as the size, shape, and arrangement of the mask elements must account for these effects to mitigate image artifacts. An in-depth discussion of how our coded aperture was tailored to the near field can be found in Frame et al. \cite{frame}.

For Compton cameras, one of the more critical design considerations is the detector granularity. This characteristic can be defined as the ability of the imager to discriminate multiple interactions induced by the same incident gamma ray. In other words, a finer granularity corresponds to a smaller volume in which two interactions can be discriminated; thereby increasing the fraction of detected events that are correctly identified. Furthermore, the energy and position resolution of the detectors must be considered as these two parameters strongly impact the attainable image resolution \cite{nurdan}.

The remainder of this article demonstrates the feasibility of imaging the $^{225}$Ac daughters in small animals via coded aperture and Compton methods. Section \ref{sec:dmi} describes the design of the imaging system, hereinafter referred to as the Dual-Modality Imager, and evaluates its performance in terms of imaging sensitivity and resolution. Section \ref{sec:results} describes our small-animal experiments and presents coded aperture and Compton images of $^{221}$Fr and $^{213}$Bi, respectively, in tumor-bearing mice injected with $^{225}$Ac agents. A discussion of these results follows in Section \ref{sec:diss}. Finally, Section \ref{sec:concl} provides closing remarks.

\section{Imaging System}
\label{sec:dmi}
\subsection{The Dual-Modality Imager}
The Dual-Modality Imager is a gamma-ray imaging platform that functions as both a coded aperture and Compton imager. This platform was \textit{not} designed for small-animal applications and is merely employed here as a proof-of-concept prototype. The Dual-Modality Imager consists of two three-dimensional (3-D) position-sensitive high-purity germanium (HPGe) double-sided strip detectors (DSSDs) manufactured by Lawrence Berkeley National Laboratory (LBNL). The two detectors are housed in the same cryostat and separated by a distance of $10$~mm to maximize the solid angle between them. Each detector consists of a planar HPGe crystal with an active volume of $74\times74\times15$~mm$^3$ surrounded by $2$-mm wide guard rings to reduce leakage current. The opposite faces of each detector have $37\times37$ orthogonal strip electrodes of $2$-mm pitch. This electrode configuration gives an intrinsic granularity of $1369$ $2\times2$-mm$^2$ pixels.

\begin{figure}[!tbp]
  \centering
  \subfloat[]{\includegraphics[width=0.501\columnwidth]{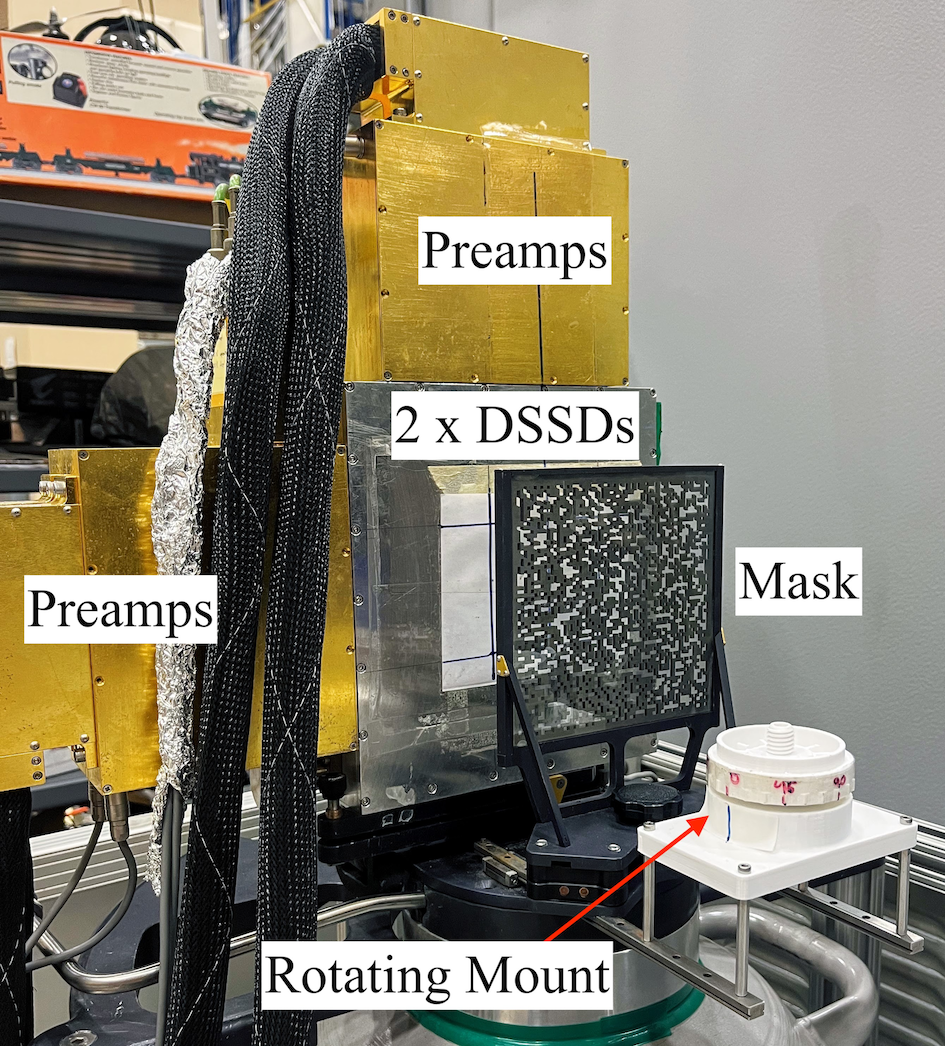}}
  \hspace{0.5cm}
  \subfloat[]{\includegraphics[width=0.43\columnwidth]{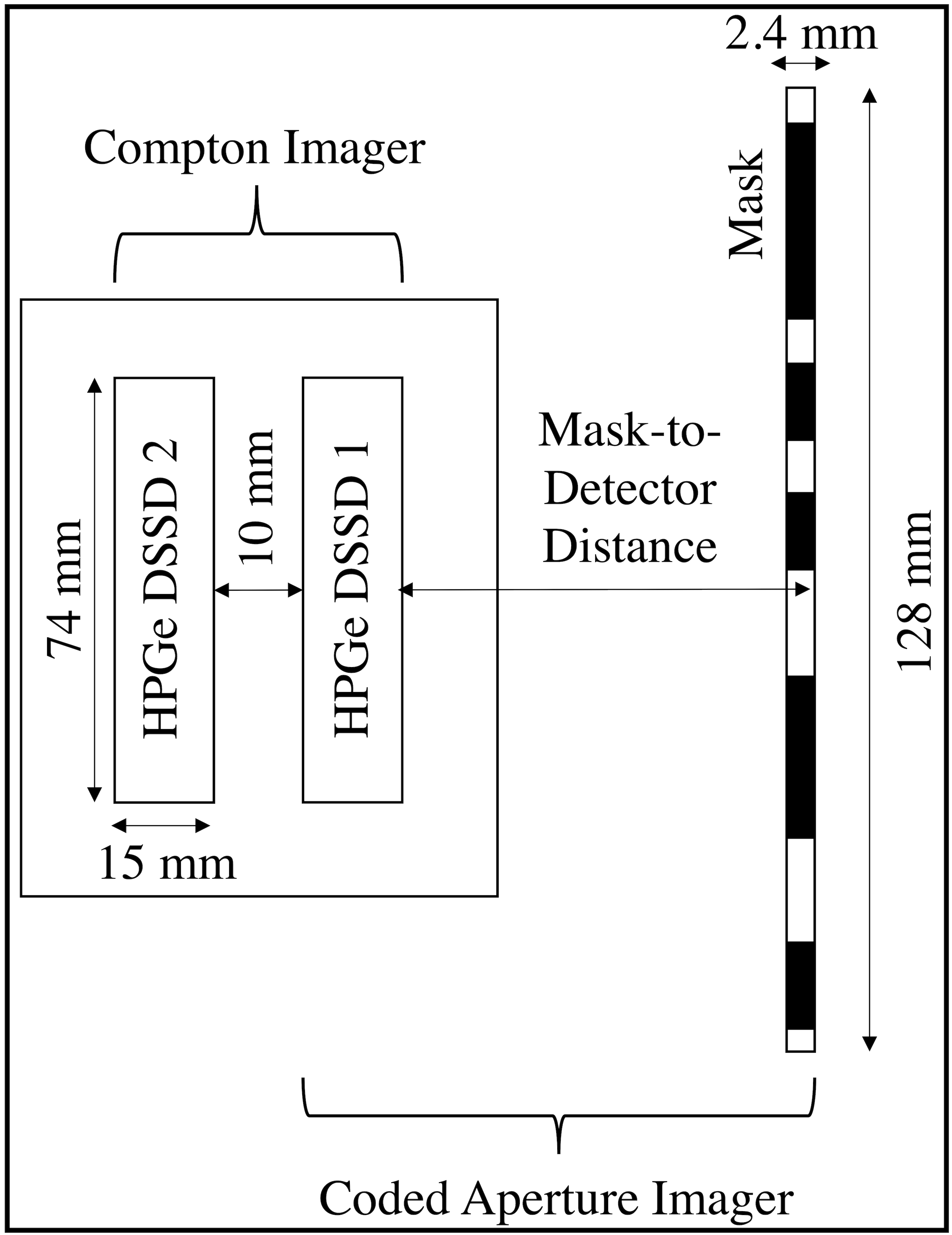}}
  \caption{(a) The Dual-Modality Imager featuring two DSSDs and a coded aperture. (b) A schematic of the coded aperture and Compton geometries \cite{frame}.}
  \label{fig:1}
\end{figure}

A custom-made coded aperture is situated at an adjustable distance in front of the first detector facing the source. Fig. \ref{fig:1}a shows a close-up of the detector and coded aperture arrangement. The mask is fabricated from $2.4$-mm-thick tungsten, which ensures roughly $90\%$ attenuation at $250$~keV, and consists of $64\times64$ square elements.  Each square element has a $2\times2$~mm$^2$ face to match the detector pixel size and limit collimation effects. The elements are collectively arranged in a random pattern with a $50\%$ open fraction. The pattern was optimized across a range of magnifications using a combinatorial search technique \cite{frame}.

The coded aperture combined with the first detector constitutes the coded aperture imager, while the first and second detectors collectively form the Compton camera as illustrated in Fig. \ref{fig:1}b. Note this geometric arrangement does not allow the simultaneous use of both modalities without a cost to imaging performance. Because the Compton modality is characterized by a finite angular resolution, decreasing the distance between the imager and source significantly improves the image spatial resolution. However, with the coded aperture in place, the minimum standoff distance is limited to the distance between the mask and front detector. Furthermore, the coded aperture allows fewer photons to reach the detector; this results in a lower Compton imaging sensitivity. For these reasons, the two modalities are operated separately with the mask removed in the Compton mode.

In designing a more specialized system for time-sensitive applications that require simultaneous use, a third detector can be introduced on the opposite side of the mask. In this configuration, the Compton camera would be formed by the original two detectors, and the coded aperture imager would be formed by the mask and newly-added third detector. This configuration would allow the source to be positioned in between the two modalities at an optimal distance that maximizes both the sensitivity and resolution of each modality.

\subsection{Imaging Sensitivity}
Imaging sensitivity can be defined as the ratio of photons selected for image reconstruction (in cps) to the activity (in MBq) of the source. To determine the sensitivity of the Dual-Modality Imager, we acquired coded aperture and Compton data from an $^{225}$Ac-filled microsphere with a well-defined activity. The microsphere has an inner volume of about $40$~$\mu$L and was filled with a mixture of water and $^{225}$Ac in secular equilibrium with $^{221}$Fr and $^{213}$Bi. The total activity was $7.0$~kBq and $6.5$~kBq at the time of the coded aperture and Compton measurements, respectively.

In the coded aperture mode, the $7.0$~kBq microsphere was positioned at a source-to-detector distance of $145$~mm in the center of the field-of-view. This standoff distance was selected to match that of the small animal in the coded aperture mode. A total of $2700$ events were acquired at the $218$~keV emission line of $^{221}$Fr after $41$~minutes. This equates to a sensitivity of about $160$ cps/MBq ($0.016$\%) for $^{221}$Fr at a source-to-detector distance of $145$~mm.

In the Compton mode, the $6.5$~kBq microsphere was positioned at a source-to-detector distance of $35$~mm in the center of the field-of-view. This standoff distance was selected to match that of the small animal in the Compton mode. A total of $4400$ events were acquired at the $440$~keV emission line of $^{213}$Bi after $75$~minutes. This equates to a sensitivity of about $150$ cps/MBq ($0.015$\%) for $^{213}$Bi at a source-to-detector distance of $35$~mm.

\subsection{Image Resolution}
Image resolution is strongly controlled at the detector level. The position resolution of the detectors impacts the image resolution of both the coded aperture and Compton modalities. For all practical purposes, the lateral position resolution of DSSDs in FWHM can be approximated as the strip pitch; our DSSDs have a $2$-mm pitch. The depth resolution of our detectors was previously reported to be as good as $0.5$~mm in FWHM \cite{chivers}.

Furthermore, the energy resolution of the detectors solely impacts the Compton image resolution. The uncertainty in the first energy deposition, of which the finite energy resolution of the detectors is a contributor, propagates to the uncertainty in the opening angle of the Compton cone. This error propagation blurs the Compton image. Our detectors exhibit excellent energy resolution, averaging between $0.2\%$ and $0.5\%$ at $662$~keV, due to the favorable detection properties of HPGe.

Accorsi \cite{accorsi} derives the following expression for the lateral resolution $\delta r_{CA}$ of a coded aperture imager in FWHM:
\begin{align}
\delta r_{CA} \approx \left(\frac{1}{m-1}\right) \sqrt{\left[\left(m\right)\left(w_m\right)\right]^2 + w_d^2} \label{eq:resca}
\end{align}
where $w_d$ is the lateral position resolution of the detectors and $w_m$ is the mask pixel width; in this work, $w_d=w_m=2$~mm. The parameter $m$ is the magnification factor defined as:
\begin{align}
  m = 1 + \frac{b}{a} \label{eq:magn}
\end{align}
where $m$ represents the ratio of the projection of an individual mask element to the mask element itself, $a$ is the normal distance between the source and mask planes, and $b$ is the normal distance between the mask and detector planes.

According to Eq. \ref{eq:resca}, given a fixed coded aperture design, the resolution improves by increasing the magnification factor $m$. This objective can be realized by decreasing the source-to-mask distance $a$ and/or increasing the mask-to-detector distance $b$. In this work, we select both $a$ and $b$ to be as small as possible to maximize sensitivity. Given the geometric constraints of the coded aperture and small-animal mounts, we are limited to $a=95$~mm and $b=50$~mm so that $m=1.5$. At this magnification, the resolution of the coded aperture imager should be about $\delta r_{CA} \approx 6.9$~mm in FWHM.

\begin{figure}[!tbp]
   \centering
   \includegraphics[width=0.7\columnwidth]{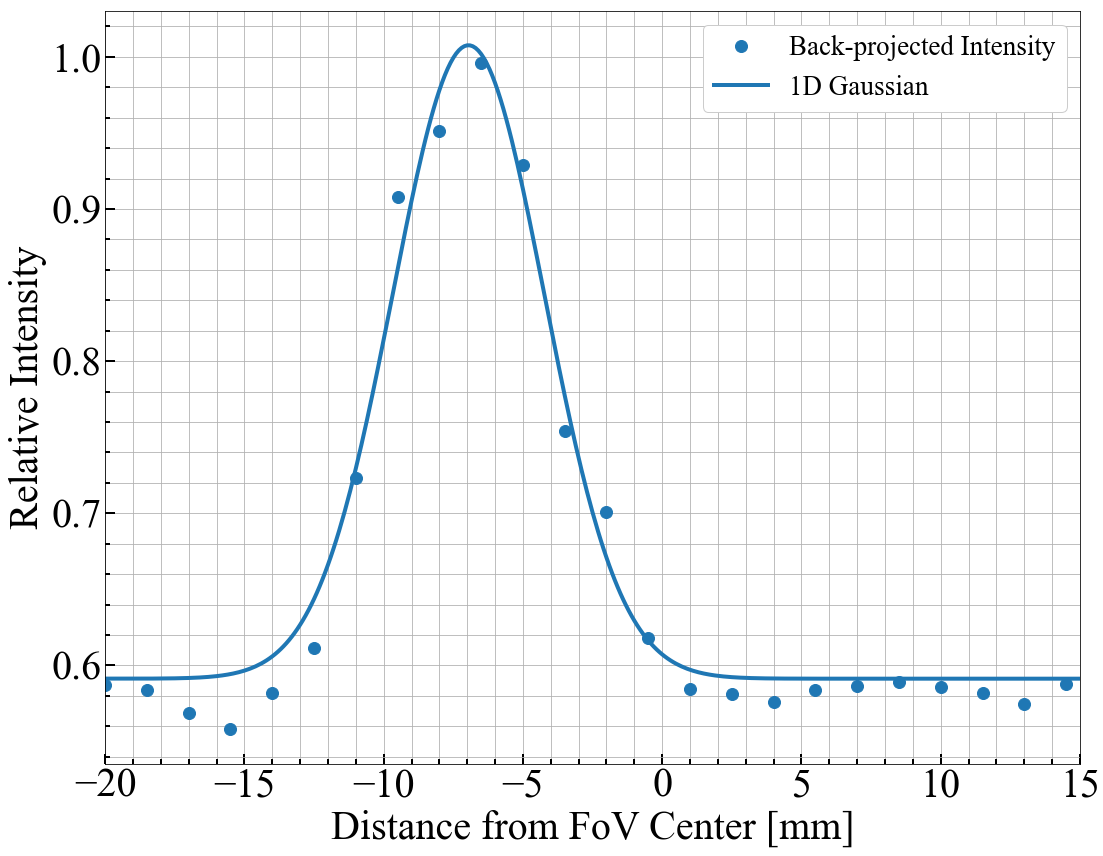}
   \caption{Linear cross section of a back-projection of a $^{57}$Co disk source of $2$-mm diameter positioned near the center of the field-of-view at a magnification of $m=1.5$ in the coded aperture mode. The cross section was taken at the highest intensity image voxel. A 1-D Gaussian model was fit to the intensity profile to provide an estimate of the image resolution in FWHM.}
   \label{fig:2}
\end{figure}

We validated the theoretical coded aperture resolution by evaluating the back-projection of a $^{57}$Co disc source in the coded aperture mode. The disc source has a $2$~mm active diameter and was positioned near the center of the field-of-view at a distance of $a=95$ mm from the mask with the mask positioned at a distance of $b=50$ mm from the detector. Fig. \ref{fig:2} shows a linear cross section of the back-projection. This cross section was taken at the highest intensity image voxel and was fitted with a one-dimensional (1-D) Gaussian model. The FWHM of the Gaussian was found to be $6.4$~mm. A 1-D deconvolution of the source diameter from the FWHM of the Gaussian gives an estimate of the resolution. The resolution was estimated to be about $6.3$~mm; this is consistent with the theoretical value of $\delta r_{CA} \approx 6.9$~mm.

The lateral resolution $\delta r_{CI}$ of a Compton imager in FWHM can be defined as:
\begin{align}
  \delta r_{CI} \approx \left( 2z \right) \tan \left( \frac{\delta\theta}{2} \right) \label{eq:resci}
\end{align}
where $z$ can be approximated as the normal distance between the source and the detector surface facing the source and $\delta\theta$ is the angular resolution. In Compton imaging, $\delta\theta$ is governed by uncertainties in the energy depositions and positions of interaction. The former uncertainty is governed by Doppler broadening and the finite energy resolution of the detectors, while the latter uncertainty is governed by the finite position resolution of the detectors.

The most straightforward approach to determine the Compton angular resolution $\delta\theta$ is to generate an angular resolution metric (ARM) distribution. The ARM metric can be defined as the angular separation between the Compton cone and known source location. We generated ARM distributions from simulated (Geant4) data at both $440$~keV and $662$~keV. We also generated an ARM distribution from experimental data at $662$~keV to demonstrate the consistency between experiment and simulations. On the experimental side, we acquired Compton data at $662$~keV from a $^{137}$Cs disc source. The source has a $2$-mm active diameter and was positioned at a distance of $200$~mm from the surface of the front detector. The standoff distance was selected to be large so that the source appeared point-like to the imager.

\begin{figure}[!tbp]
  \centering
  \subfloat[]{\includegraphics[width=0.7\columnwidth]{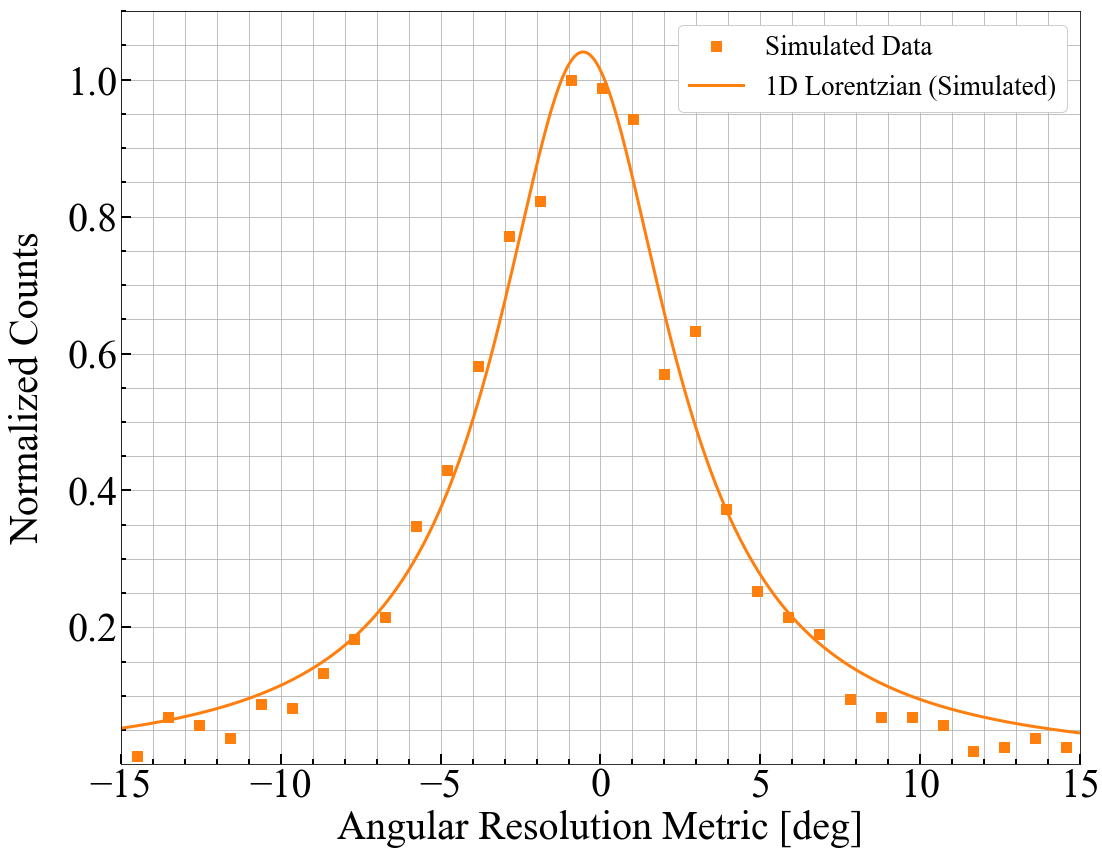}}
  \\
  \subfloat[]{\includegraphics[width=0.7\columnwidth]{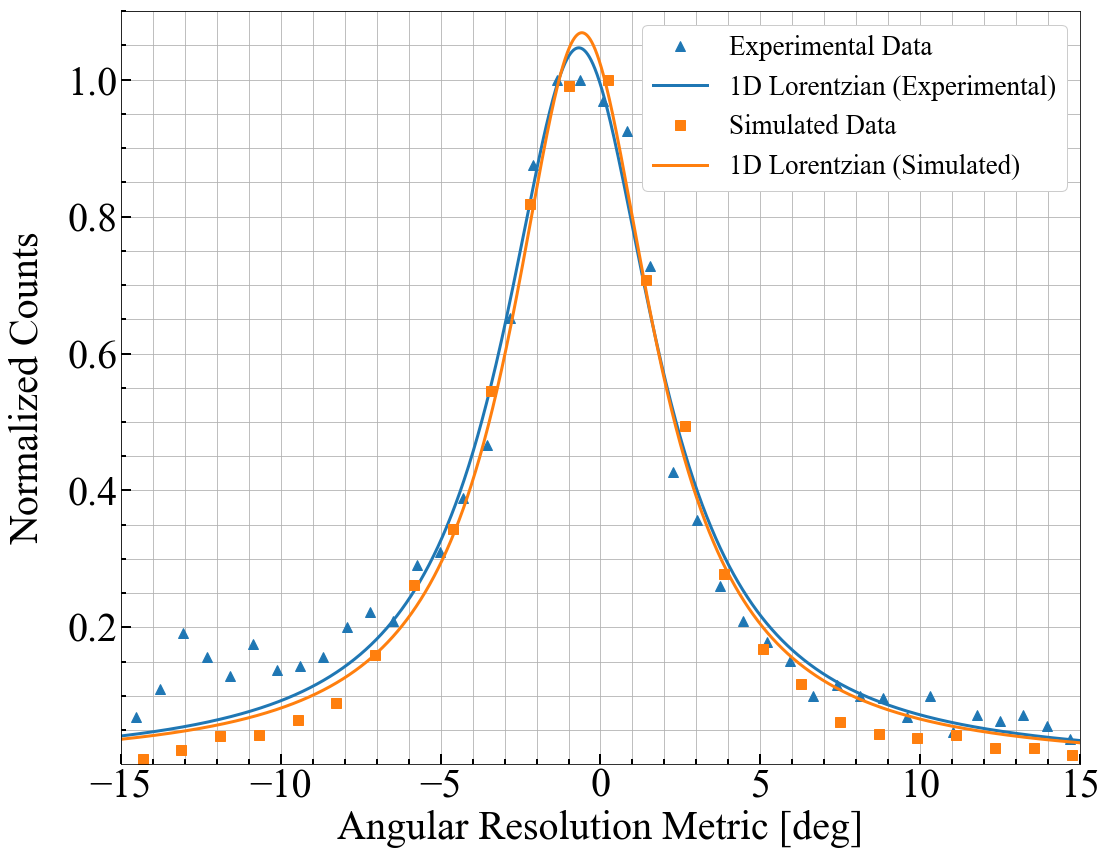}}
  \caption{ARM distribution at (a) $440$~keV from simulated data only and (b) $662$~keV from both simulated and experimental data. A 1-D Lorentzian model was fit to each distribution to provide an estimate of the angular resolution in FWHM.}
  \label{fig:3}
\end{figure}

Fig. \ref{fig:3} shows the ARM distributions fitted with a 1-D Lorentzian model. The FWHM of the Lorentzian provides an estimate of the angular resolution $\delta\theta$. The simulated and experimental ARM distributions at $662$~keV are consistent, showing FWHMs of $5.4^\circ$ and $5.5^\circ$, respectively. Thus, we can have confidence in the simulated ARM distribution at $440$~keV, which exhibits a FWHM of $6.7^\circ$. The improvement in angular resolution with increased energy should be expected. At higher energies, Doppler broadening and energy resolution have a smaller impact, and photons produce longer lever arms due to a greater penetrative power. All of the above contribute to a more accurate Compton cone.

According to Eq. \ref{eq:resci}, angular errors are magnified by the source-to-detector distance $z$. For this reason, we want to position the source as close as possible to the detector; this is also advantageous from a sensitivity standpoint. Given the geometric constraints of the detector and small-animal containments, the minimum allowable source-to-detector distance is $z = 35$~mm. At this distance for a $440$~keV source, the intrinsic resolution of the Compton imager should be about $\delta r_{CI} \approx 4.1$~mm in FWHM.

\section{Experiments and Results}
\label{sec:results}
\subsection{Phantom Filled with $^{225}$Ac}
\label{subsec:phantom}
\begin{figure}[!tbp]
  \centering
  \subfloat[]{\includegraphics[width=0.454\columnwidth]{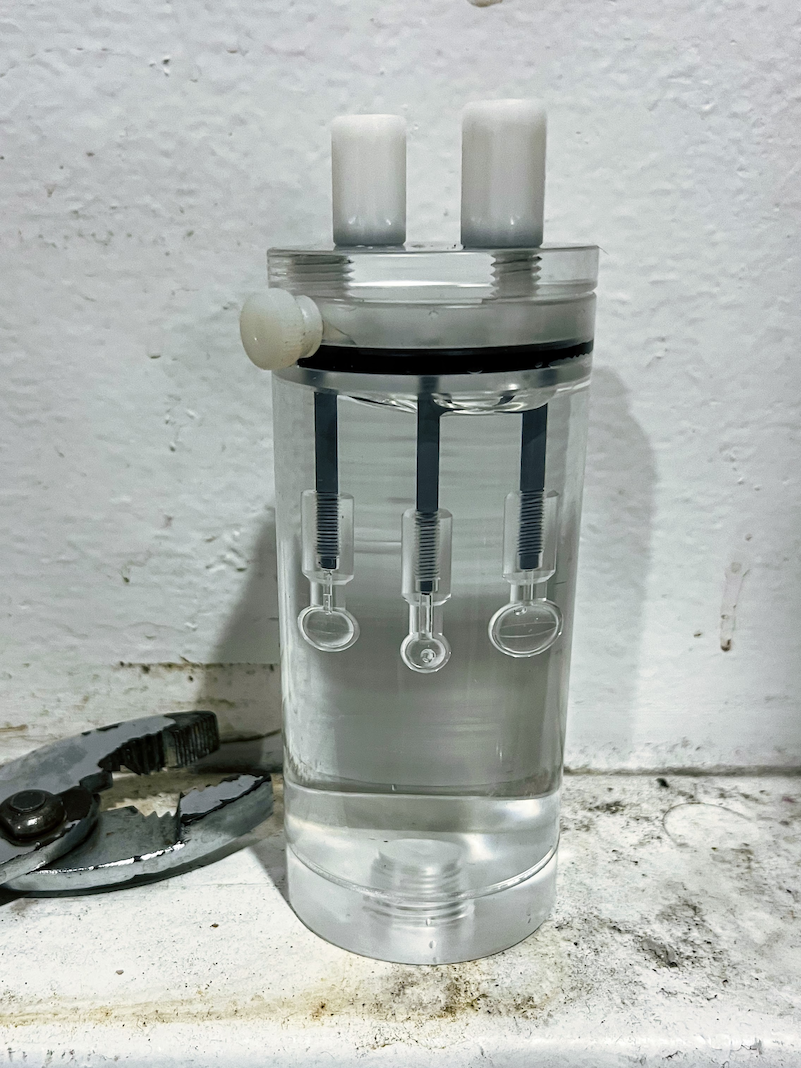}}
  \hspace{0.5cm}
  \subfloat[]{\includegraphics[width=0.3485\columnwidth]{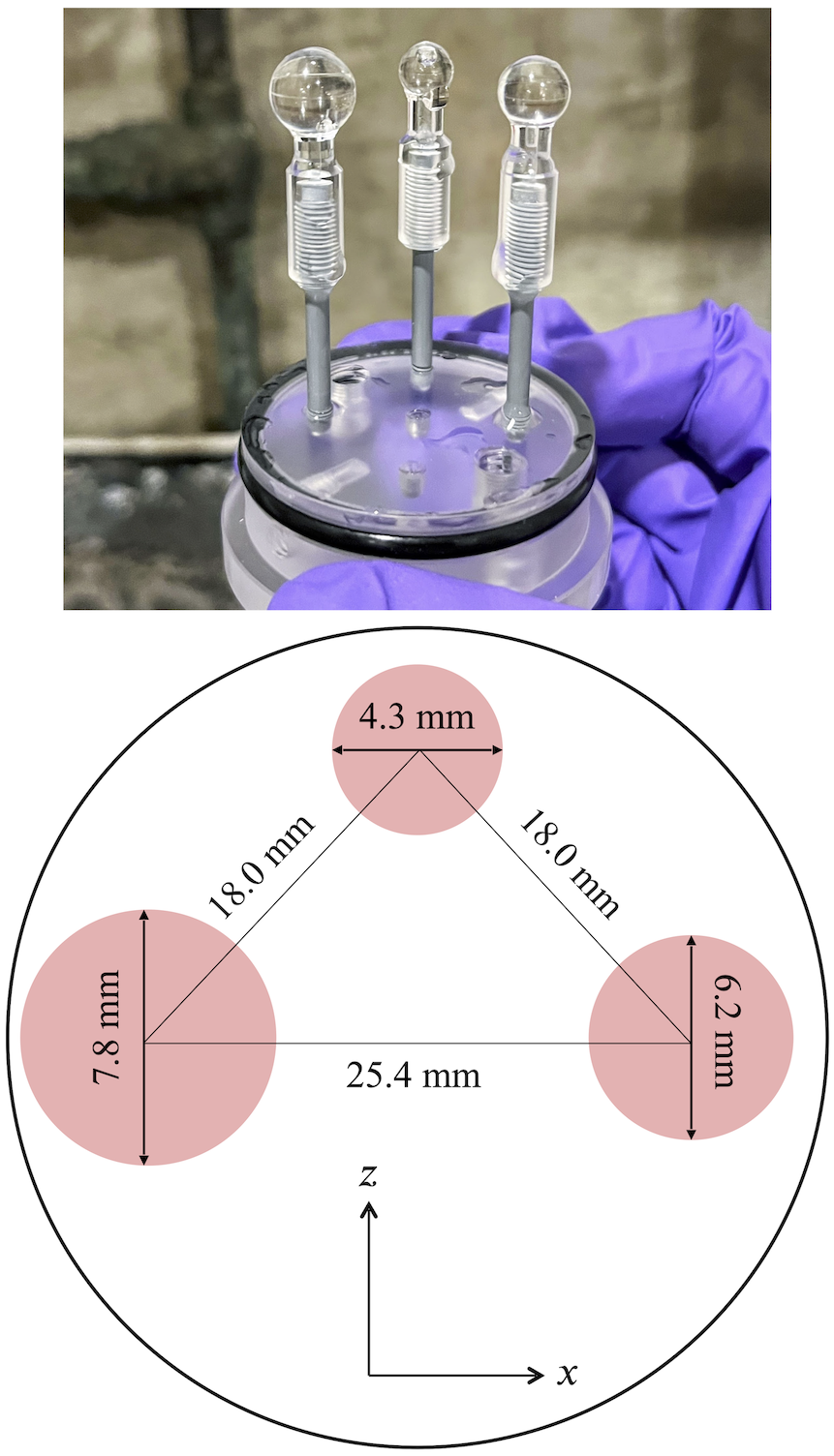}} \\
  \subfloat[]{\includegraphics[width=0.4183\columnwidth]{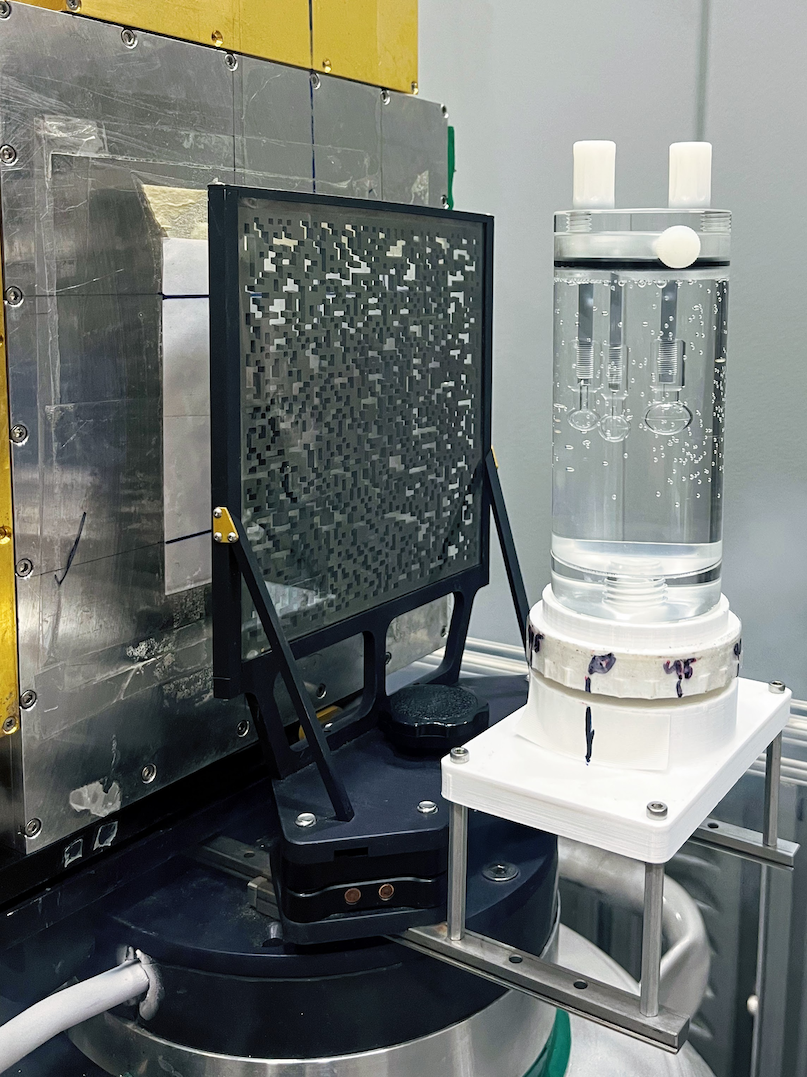}}
  \hspace{0.5cm}
  \subfloat[]{\includegraphics[width=0.3425\columnwidth]{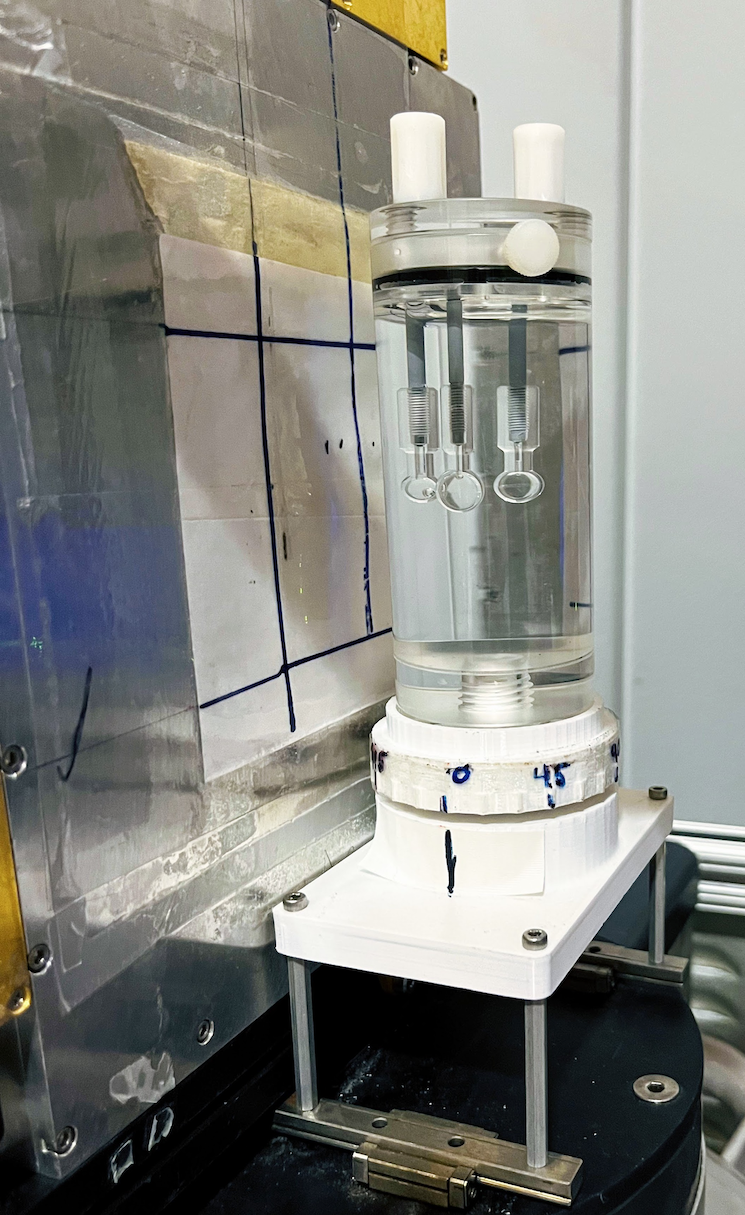}}
  \caption{(a) Cylindrical body of phantom. (b) Phantom contains three micro-hollow spheres with inner diameters of $4.3$~mm, $6.2$~mm, and $7.8$~mm. The sphere centers are positioned in a triangular fashion on the same $(x, z)$ transverse plane. (c) Experimental setup of the phantom in the coded aperture mode at a source-to-detector distance of $145$~mm with a mask-to-detector distance of $50$~mm and (d) Compton mode at a source-to-detector distance of $55$~mm.}
  \label{fig:4}
\end{figure}

This section presents a coded aperture image of $^{221}$Fr and Compton image of $^{213}$Bi in an $^{225}$Ac-filled phantom. These images provide quantification factors for the subsequent small-animal images. Fig. \ref{fig:4}a shows the phantom, which has a cylindrical body with an inner diameter of $40$~mm and inner height of $82$~mm. The body houses three micro-hollow spheres of various sizes. The inner diameters of the spheres are about $4.3$~mm, $6.2$~mm, and $7.8$~mm, and the corresponding inner volumes are about $40$~$\mu$L, $125$~$\mu$L, and $250$~$\mu$L, respectively.

Fig. \ref{fig:4}b shows the configuration of the spheres inside the phantom. The sphere centers are positioned in a triangular fashion on the same $(x, z)$ transverse plane. The centers of the large and medium spheres are separated by a distance of $25.4$~mm. The center of the small sphere is located at a distance of $18.0$~mm from the centers of both the large and medium spheres.

The phantom body was filled completely with water. The spheres were each filled fully with the same mixture of water and $^{225}$Ac in secular equilibrium with $^{221}$Fr and $^{213}$Bi. The activity concentration was $0.53$~kBq/$\mu$L at the time of preparation. Given this concentration, the smallest sphere contained $22$~kBq, the medium sphere contained $67$~kBq, and the largest sphere contained $133$~kBq.

Fig. \ref{fig:4}c shows the experimental setup of the phantom in the coded aperture mode at $2$~days post-preparation. By the time of this measurement, the activities inside the spheres had decayed to $19$~kBq, $58$~kBq, and $117$~kBq in order of the smallest to largest sphere. The phantom was centrally positioned on top of a rotating mount. The rotation axis was located at a distance of $95$~mm from the mask with the mask positioned at a distance of $50$ mm from the detector. In this configuration, the resolution of the coded aperture imager theoretically should be about $6.9$~mm in FWHM. The mount was rotated in $45^\circ$ increments. At each viewing angle, $3.1 \times 10^{4}$ events were acquired on average at the $218$-keV emission line of the daughter $^{221}$Fr after $30$~minutes. The total imaging time amounted to $4$~hours.

Fig. \ref{fig:4}d shows the experimental setup of the phantom in the Compton mode at $14$~days post-preparation. By the time of this measurement, the activities inside the spheres had decayed to $8$~kBq, $25$~kBq, and $51$~kBq in order of the smallest to largest sphere. The phantom was centrally positioned on top of a rotating mount. For maximum resolution and sensitivity, the mount was positioned as close as possible to the detector cryostat with the rotation axis located at a distance of $55$~mm from the surface of the first detector. In this configuration, the resolution of the Compton imager theoretically should be about $6.4$~mm in FWHM at $440$~keV. The mount was rotated in $45^\circ$ increments. At each viewing angle, $8.4 \times 10^{3}$ events were acquired on average at the $440$-keV emission line of the daughter $^{213}$Bi after $30$~minutes. The total imaging time amounted to $4$~hours.

The restriction of the imager location to eight positions (associated with $45^\circ$ source rotations) is significantly less than used in standard SPECT imaging. Fewer rotations are required, because both the coded aperture and Compton imaging modalities provide parallax, and thus depth information from a single viewing angle. This is not possible with a standard gamma camera.

\begin{figure}[!htb]
  \centering
  \subfloat[]{\includegraphics[width=0.499\columnwidth]{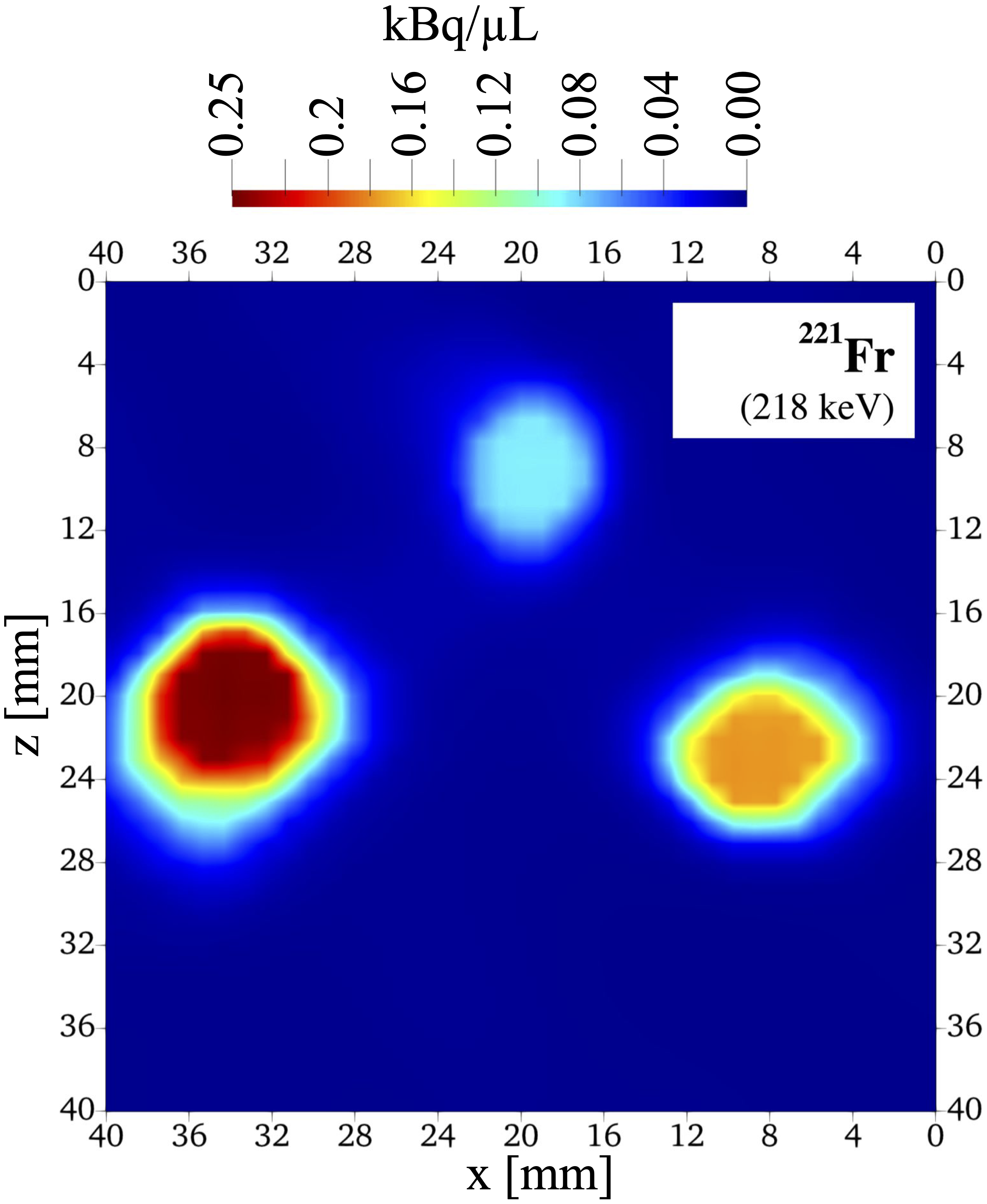}}
  \hfill
  \subfloat[]{\includegraphics[width=0.499\columnwidth]{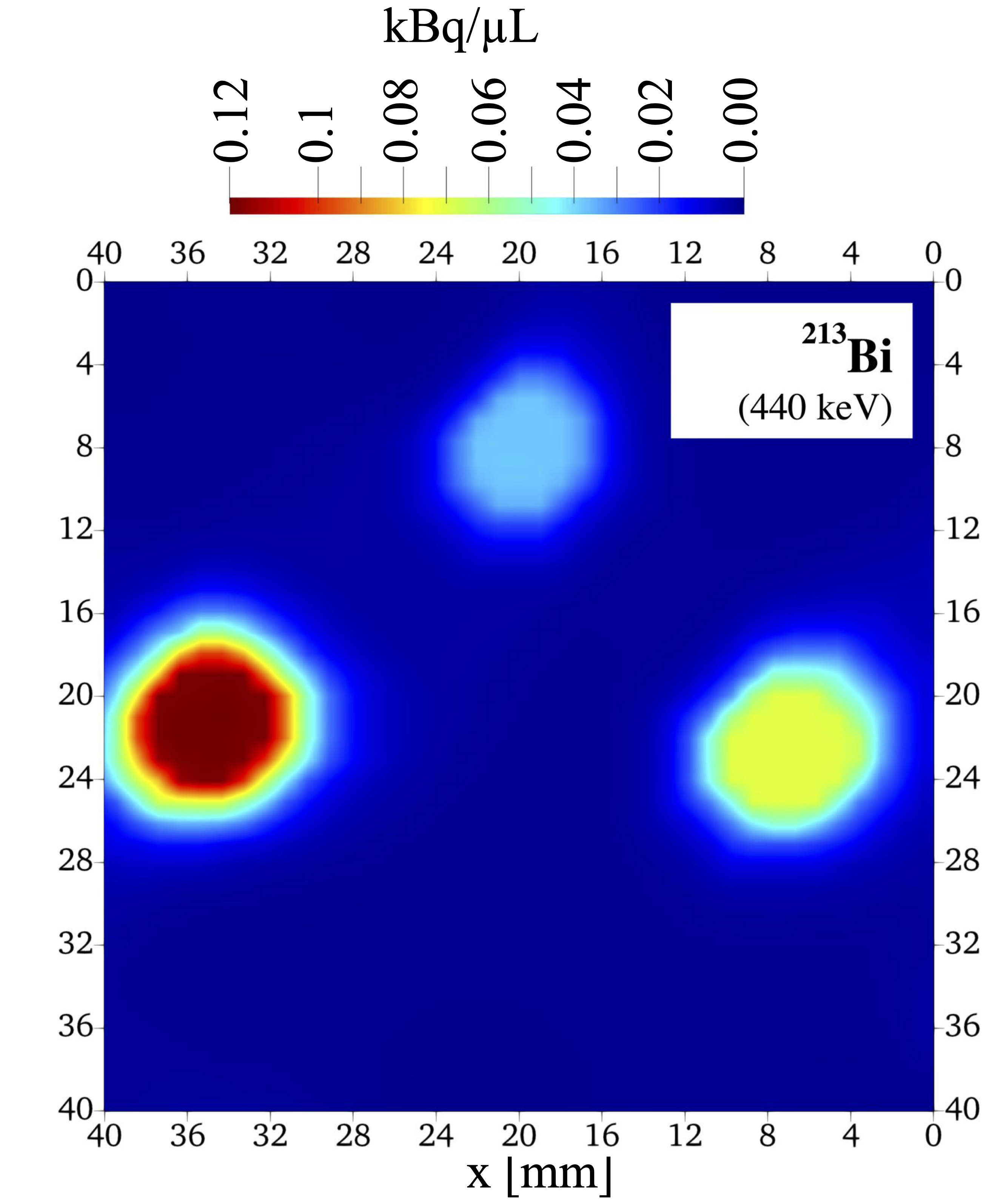}}
  \caption{(a) Coded aperture and (b) Compton images of the $^{225}$Ac-filled phantom displayed as two-dimensional (2-D) transverse $(x,z)$ slices. The coded aperture image was generated from a total of $2.5 \times 10^5$ events at $218$~keV, and the Compton image was generated from a total of $6.7 \times 10^4$ events at $440$~keV.}
  \label{fig:5}
\end{figure}

Using maximum-likelihood expectation-maximization (ML-EM) with total variation (TV) as described by Frame \cite{frame2}, the phantom was reconstructed in 3-D from the coded aperture and Compton data at the $218$-keV and $440$-keV emission lines, respectively. Both the coded aperture and Compton images were generated from all eight projections after $250$~iterations. Fig. \ref{fig:5} shows the central $(x, z)$ transverse slices of the phantom images.

The image intensity scales are in units of activity per volume (kBq/$\mu$L). To get activity units, respective quantification factors $Q$ were derived from the coded aperture and Compton images of the phantom:
\begin{align}
  Q = \frac{I}{\left(A\right) \left(T\right) \left(\frac{1}{R^2}\right) \exp\left[-\left(\mu_{H2O}\right)\left(X\right)\right]}
  \label{eq:quantfact}
\end{align}
where $I$ is the sum total of intensity in the three reconstructed microspheres as determined by the ML-EM algorithm, $A$ is the known activity of the three microspheres at the time of measurement, $T$ is the imaging time, $R$ is the normal distance between the center of the phantom and surface of the first detector, and the exponential $\exp\left[-\left(\mu_{H2O}\right)\left(X\right)\right]$ corrects for photon attenuation in water. The attenuation distance $X$ is assumed to be the radius of the phantom body, and the water linear attenuation coefficient $\mu_{H2O}$ is evaluated at $218$~keV for the coded aperture modality and $440$~keV for the Compton modality. The coded aperture and Compton quantification factors $Q$ are assumed constant and are applied to the subsequent small-animal images of $^{221}$Fr and $^{213}$Bi, respectively. By simply plugging $Q$ into a rearranged Eq. \ref{eq:quantfact}, the activity $A$, now considered an unknown, can be determined.

Note the intensity scale of the coded aperture image is almost a factor of two higher than that of the Compton image as the former measurement took place $12$~days prior to the latter measurement. Furthermore, the coded aperture and Compton images show slightly different gamma-ray distributions. This variation can best be explained by the distinct resolution profiles of the two modalities.

\begin{table}[!htb]
  \begin{center}
  \captionsetup[subfloat]{labelformat=empty}
  \caption{Activities and diameters of the coded aperture and Compton reconstructed microspheres in Fig. \ref{fig:5} versus the ground truth. Diameters are reported as FWHM.}
  \label{tab:1}
  \subfloat[Coded Aperture (Fig. \ref{fig:5}a)]{
  \begin{tabular}{lllll}
  \toprule \cr
  True Diameter & Recon. Diameter & True Activity & Recon. Activity \cr
  [mm] & [mm] & [kBq] & [kBq] \cr
  \midrule
  4.3 & 7.2 & 19 & 19 \cr
  6.2 & 7.9  & 58 & 56 \cr
  7.8  & 8.1 & 117 & 119 \cr
  \bottomrule
  \end{tabular}
  } \\
  \subfloat[Compton (Fig. \ref{fig:5}b)]{
  \begin{tabular}{lllll}
  \toprule \cr
  True Diameter & Recon. Diameter & True Activity & Recon. Activity \cr
  [mm] & [mm] & [kBq] & [kBq] \cr
  \midrule
  4.3 & 6.8 & 8 & 9 \cr
  6.2 & 7.6 & 25 & 25 \cr
  7.8 & 7.9 & 51 & 50 \cr
  \bottomrule
  \end{tabular}
  }
  \end{center}
\end{table}

Table \ref{tab:1} summarizes the activities and diameters of the three reconstructed microspheres in Fig. \ref{fig:5}. In the coded aperture image, the sum totals of activity in each reconstructed sphere are $19$~kBq, $56$~kBq, and $119$~kBq in order of the smallest to largest sphere, respectively. These values are consistent with the actual activities at the time of measurement. Furthermore, the extents of the reconstructed spheres at the central linear cross-section are $7.2$~mm, $7.9$~mm, and $8.1$~mm in FWHM in order of the smallest to largest sphere, respectively.

In the Compton image, the sum totals of activity in each reconstructed sphere are $9$~kBq, $25$~kBq, and $50$~kBq in order of the smallest to largest sphere, respectively. These values are consistent with the actual activities at the time of measurement. Furthermore, the extents of the reconstructed spheres at the central linear cross-section are $6.8$~mm, $7.6$~mm, and $7.9$~mm in FWHM in order of the smallest to largest sphere, respectively.

The phantom images in Fig. \ref{fig:5} are representative of results that are attainable with our current imaging system. In reality, the three microspheres have an equal activity concentration, but different sizes. Given an ideal system, the reconstructed spheres would display equal intensity with the appropriate sizes for each sphere. Because of the limited spatial resolution of the Dual-Modality Imager, the three reconstructed spheres appear to be roughly the same size. Nonetheless, the reconstruction produces three spheres with the correct total activity, but distributed over the larger volumes. Consequently, the smaller two spheres appear less intense, rather than smaller, in the images. Better spatial resolution would solve this problem.

\subsection{Tumor-Bearing Mice Injected with $^{225}$Ac}
Five- to six-week-old male athymic mice were implanted subcutaneously with $5\times10^{6}$ 22Rv1 prostate cancer cells into the right flank at the Molecular Imaging Laboratory at the University of California San Francisco (UCSF). Approximately $3-5$ weeks after tumor implantation, the mice were injected with either $^{225}$Ac-Macropa-PEG8(7)-YS5 or $^{225}$Ac-DOTA-YS5 via the tail vein and sacrificed at different time intervals to evaluate the tumor-targeting specificity. The synthesis and biodistribution results for each of these $^{225}$Ac agents have been reported published separately \cite{bobba, bidkar}. Furthermore, these radioimmunotherapy agents are similar to previously reported $^{89}$Zr-labeled PET radiopharmaceuticals that target the antigen CD46, which is highly expressed on the surface of prostate and other cancers \cite{wang2021}.

We received two of the mice, hereinafter labeled as A and B, from the UCSF study. Mice A and B were injected with $20$~kBq of $^{225}$Ac-Macropa-PEG8(7)-YS5 and $^{225}$Ac-DOTA-YS5, respectively. Following injection, mice A and B were euthanized at $2$ and $4$~days, respectively. The mice were each housed in a $50$~mL falcon tube and stored in a freezer at $-7^\circ$C when not in use.

The two mice were each imaged in the coded aperture and Compton modes to determine the feasibility of using these modalities to assess the daughter redistribution of $^{225}$Ac in small animals. The coded aperture and Compton images are overlaid with CT scans. CT imaging was performed separately from the gamma-ray measurements using a MILabs U-CT system; thus, the co-registration is not precise. The following describes the experiments and results in more detail.

\begin{figure}[!tbp]
  \centering
  \subfloat[]{\includegraphics[width=0.49\columnwidth]{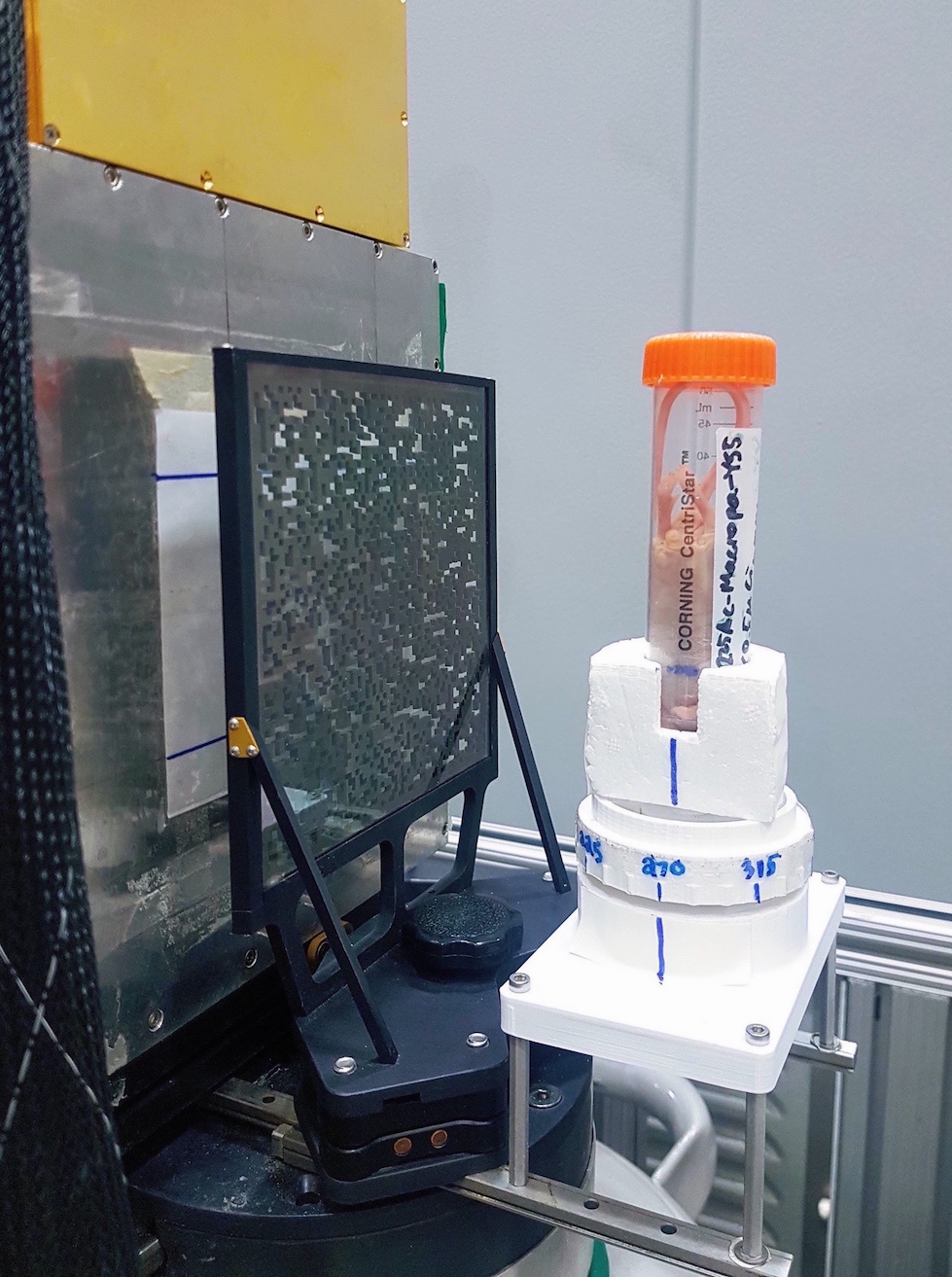}}
  \hfill
  \subfloat[]{\includegraphics[width=0.49\columnwidth]{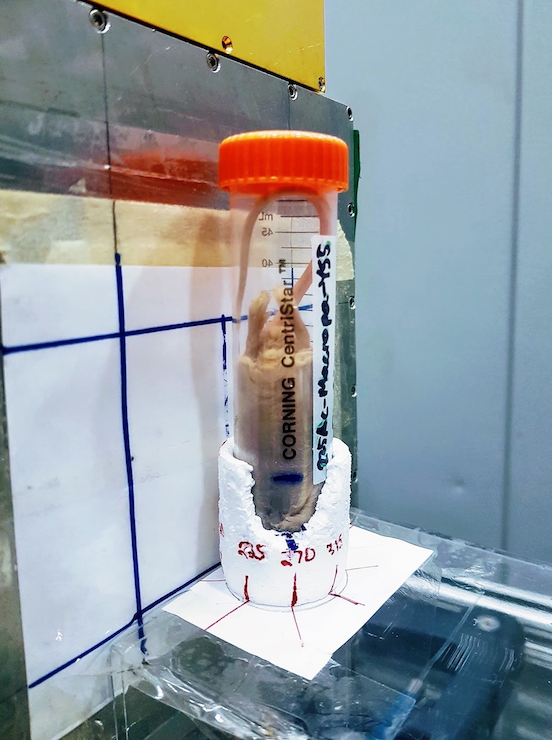}}
  \caption{Experimental setup of mouse A in the (a) coded aperture mode at a source-to-detector distance of $145$~mm with a mask-to-detector distance of $50$~mm and (b) Compton mode at a source-to-detector distance of $35$~mm.}
  \label{fig:6}
\end{figure}

Fig. \ref{fig:6}a shows the experimental setup of mouse A in the coded aperture mode at $2$~days post-injection. The falcon tube containing mouse A was centrally positioned on top of a rotating mount in an upright orientation. The rotation axis was located at a distance of $95$~mm from the mask with the mask positioned at a distance of $50$~mm from the detector. In this configuration, the resolution of the coded aperture imager theoretically should be about $6.9$~mm in FWHM. The mount was rotated in $45^\circ$ increments. At each rotation, $4.2 \times 10^3$ events were acquired on average at the $218$-keV emission line of $^{221}$Fr after $60$~minutes. The total imaging time amounted to $8$~hours.

Fig. \ref{fig:6}b shows mouse A in the Compton mode at $4$~days post-injection. To maximize the imaging sensitivity and resolution, the falcon tube was positioned as close as possible to the detector in an upright orientation. The central axis of the tube was located at a distance of $35$~mm from the surface of the first detector. In this configuration, the resolution of the Compton imager theoretically should be about $4.1$~mm in FWHM. The falcon tube was rotated by hand in $45^\circ$ increments. At each rotation, $4.1 \times 10^3$ events were acquired on average at the $440$-keV emission line of $^{213}$Bi after $75$~minutes. The total imaging time amounted to $10$~hours.

In the case of mouse B, the coded aperture and Compton experimental setups were identical to those shown in Fig. \ref{fig:6}. The coded aperture measurement of mouse B took place at $4$~days post-injection, and data were acquired for $120$~minutes at each $45^\circ$ viewing angle. The total imaging time amounted to $16$~hours, and $3.9 \times 10^3$ events were acquired on average at the $218$-keV emission line of $^{221}$Fr at each rotation. The Compton measurement of mouse B took place at $6$~days post-injection, and data were acquired for $70$~minutes at each $45^\circ$ viewing angle. The total imaging time amounted to $9.3$~hours, and $1.6 \times 10^3$ events were acquired on average at the $440$-keV emission line of $^{213}$Bi at each rotation.

Note mouse B required double the imaging time of mouse A to have sufficient statistics for coded aperture imaging. Despite mice A and B being administered the same amount of activity, the count rate observed in mouse B was about a factor of two lower. This should be expected. Mouse B was sacrificed two days later than mouse A; thereby excreting more of the radiopharmaceutical.

\begin{figure}[!tbp]
  \centering
  \captionsetup[subfloat]{justification=justified, singlelinecheck=false}
  \rotatebox{90}{\ \footnotesize$^{225}$Ac-Macropa-PEG8(7)-YS5}
  \subfloat[MIP CT]{\includegraphics[width=0.165\columnwidth]{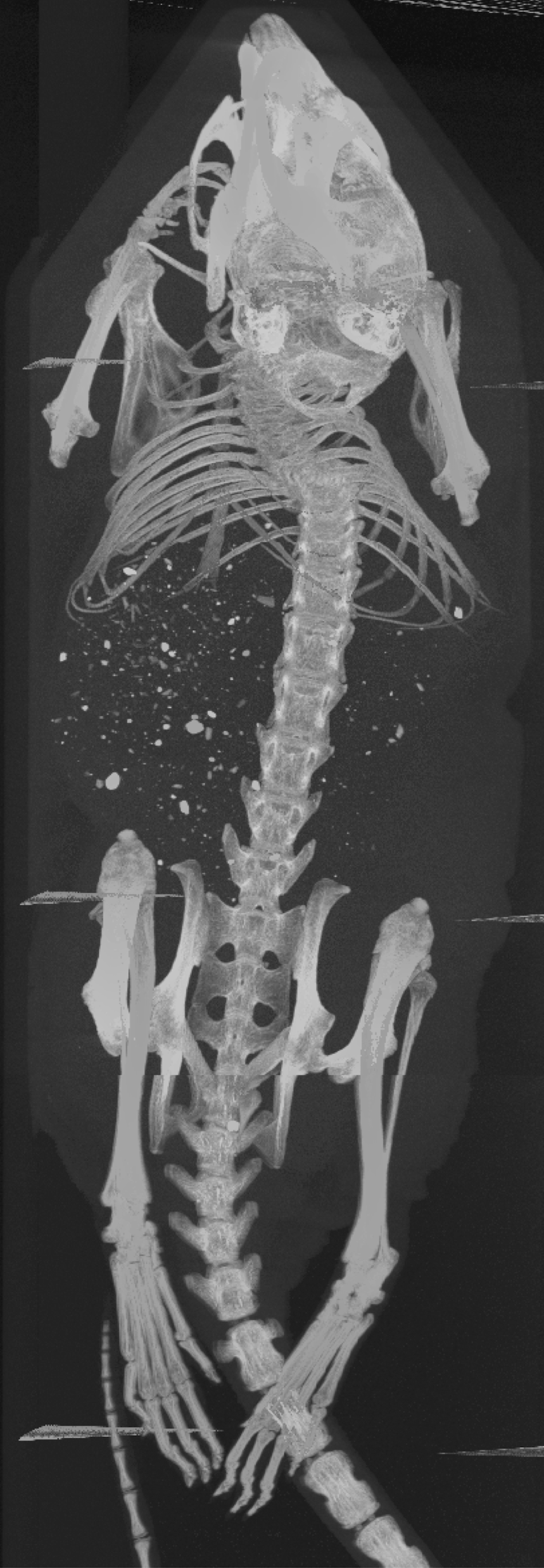}}
  \hspace{0.5cm}
  \subfloat[MIP CT + CA]{\includegraphics[width=0.2794\columnwidth]{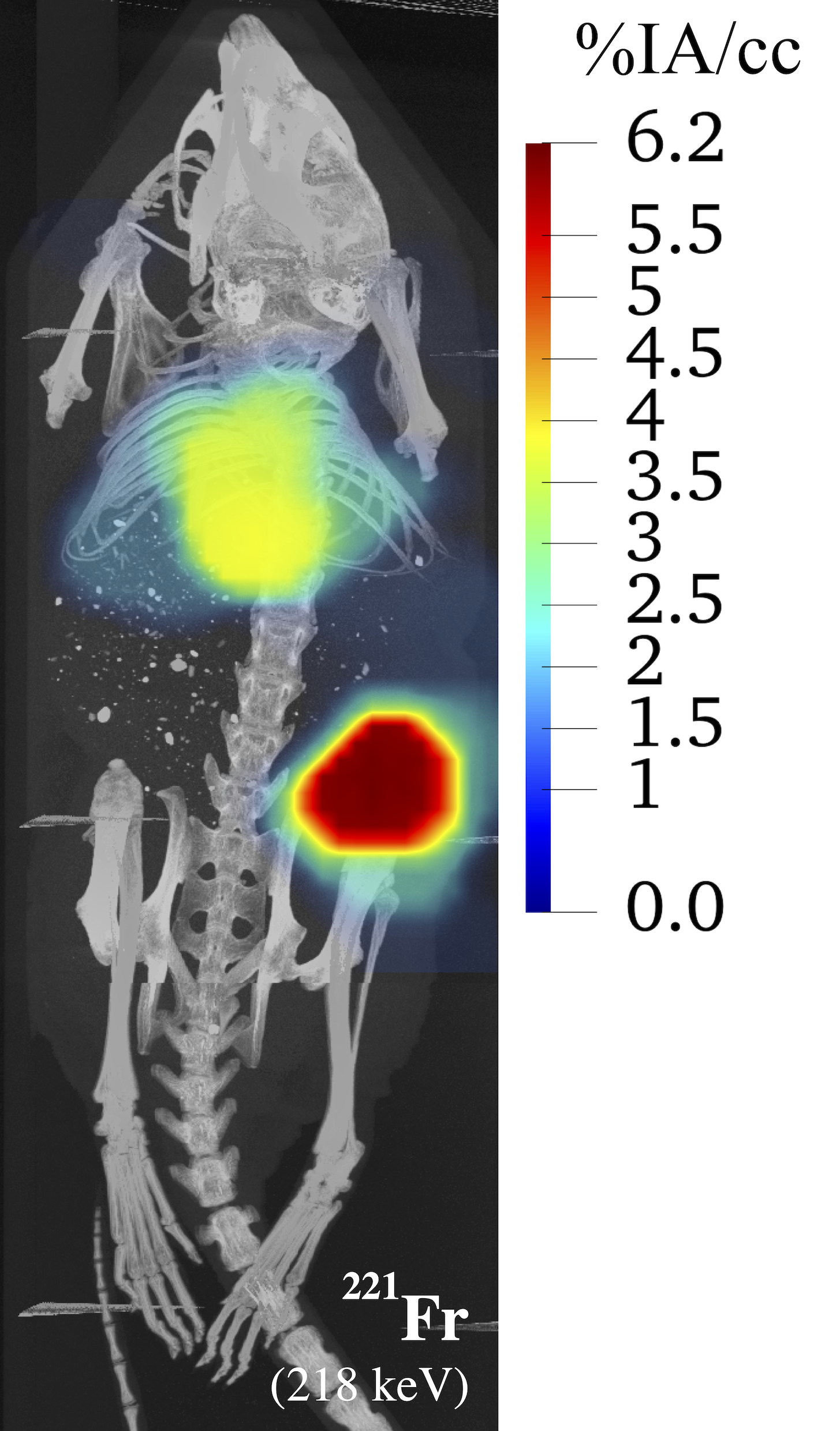}}
  \hspace{0.2cm}
  \subfloat[MIP CT + CI]{\includegraphics[width=0.2782\columnwidth]{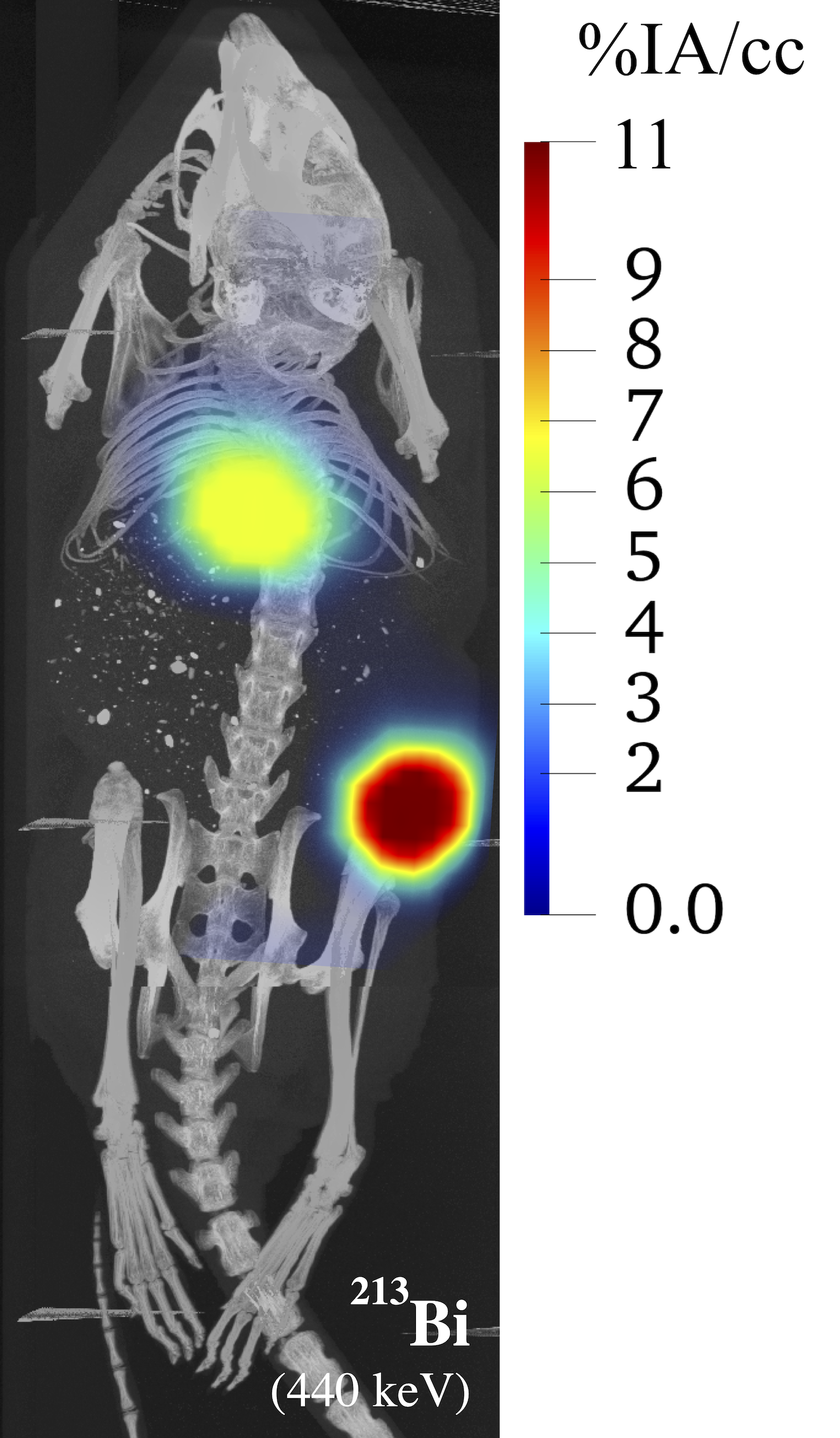}} \\
  \vspace{0.2cm}
  \rotatebox{90}{\ \footnotesize$^{225}$Ac-Macropa-PEG8(7)-YS5}
  \subfloat[2-D CT]{\includegraphics[width=0.1651\columnwidth]{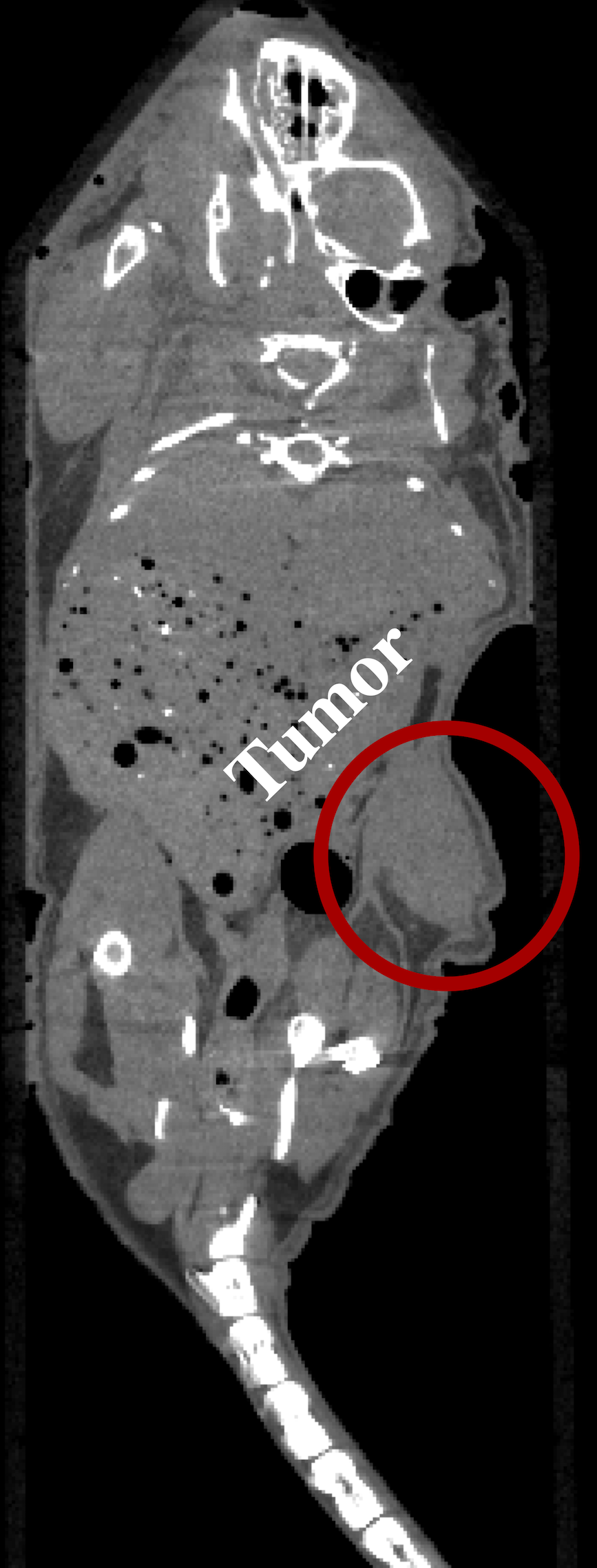}}
  \hspace{0.5cm}
  \subfloat[2-D CT + CA]{\includegraphics[width=0.2802\columnwidth]{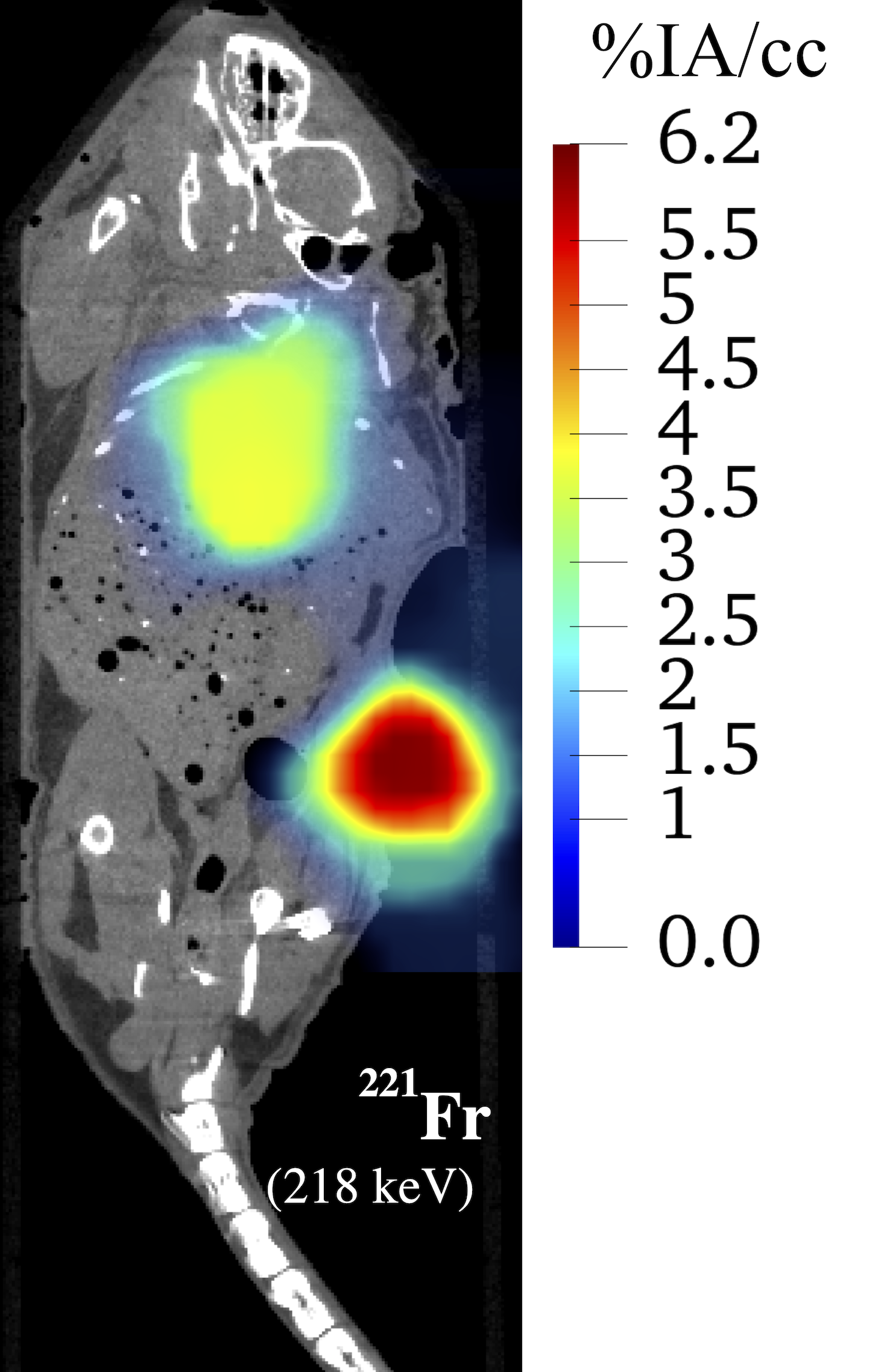}}
  \hspace{0.2cm}
  \subfloat[2-D CT + CI]{\includegraphics[width=0.275\columnwidth]{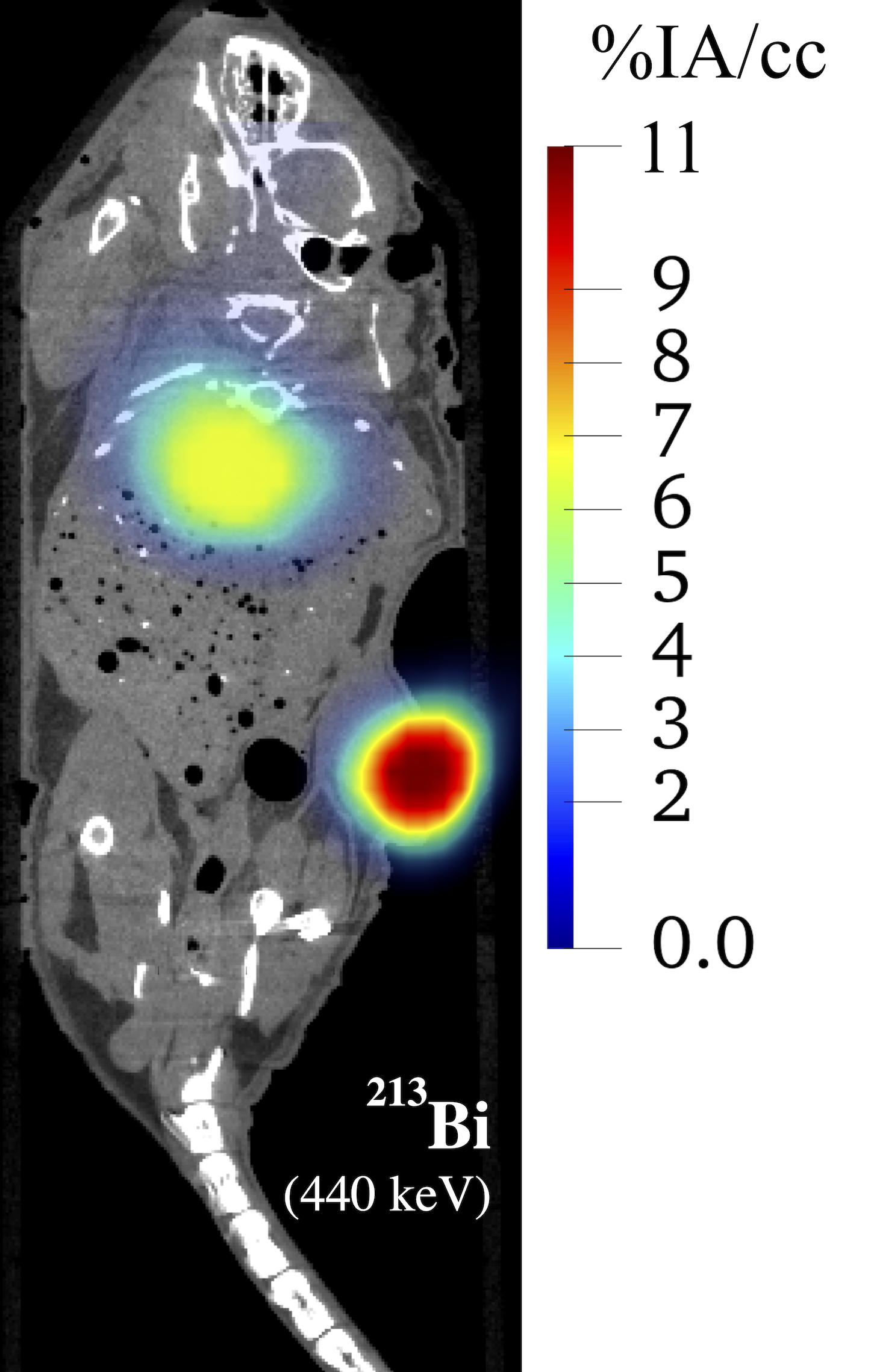}}
  \caption{(a) CT MIP of mouse A fused with the (b) coded aperture (CA) and (c) Compton (CI) MIPs. (d) CT coronal slice of mouse A fused with coronal slices of the (e) CA and (f) CI images. The CA image was generated from a total of $3.4 \times 10^4$ events at $218$~keV, and the CI image was generated from a total of $3.3 \times 10^4$ events at $440$~keV. The intensity scales are decay corrected to the day of injection.}
  \label{fig:7}
\end{figure}

\begin{figure}[!tbp]
  \centering
  \captionsetup[subfloat]{justification=justified, singlelinecheck=false}
  \rotatebox{90}{\quad\quad\footnotesize$^{225}$Ac-DOTA-YS5}
  \subfloat[MIP CT]{\includegraphics[width=0.1681\columnwidth]{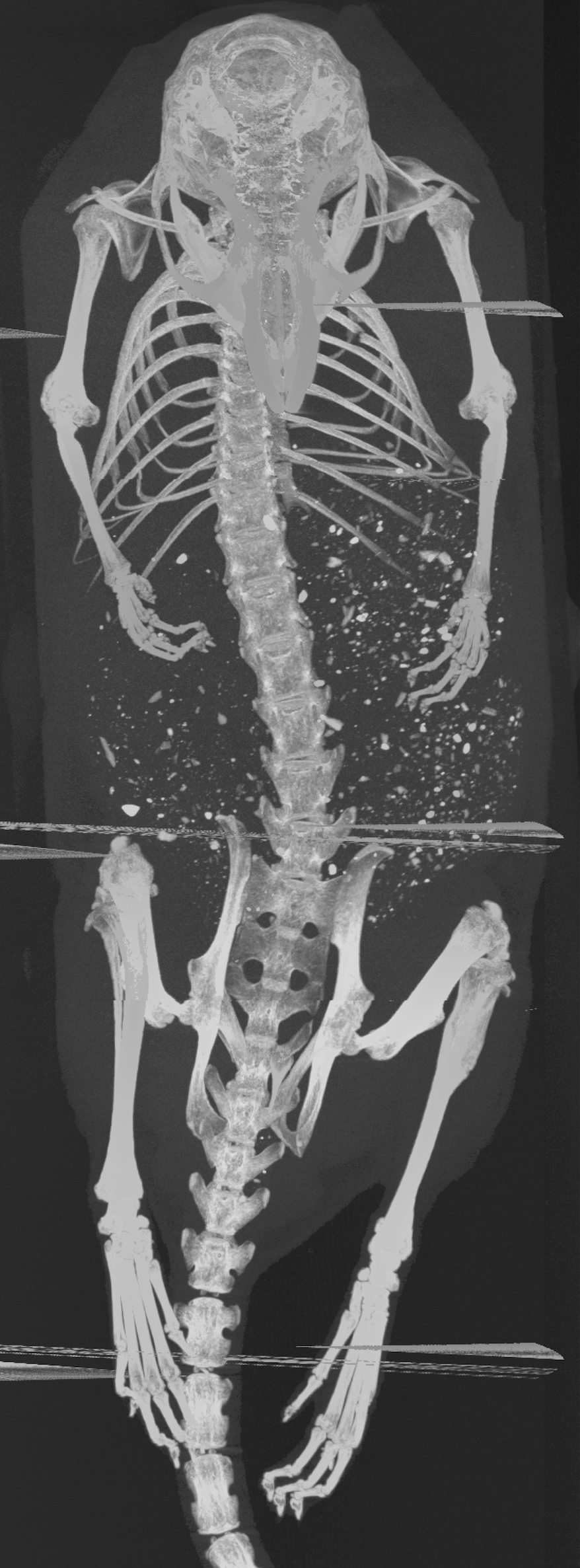}}
  \hspace{0.5cm}
  \subfloat[MIP CT + CA]{\includegraphics[width=0.28\columnwidth]{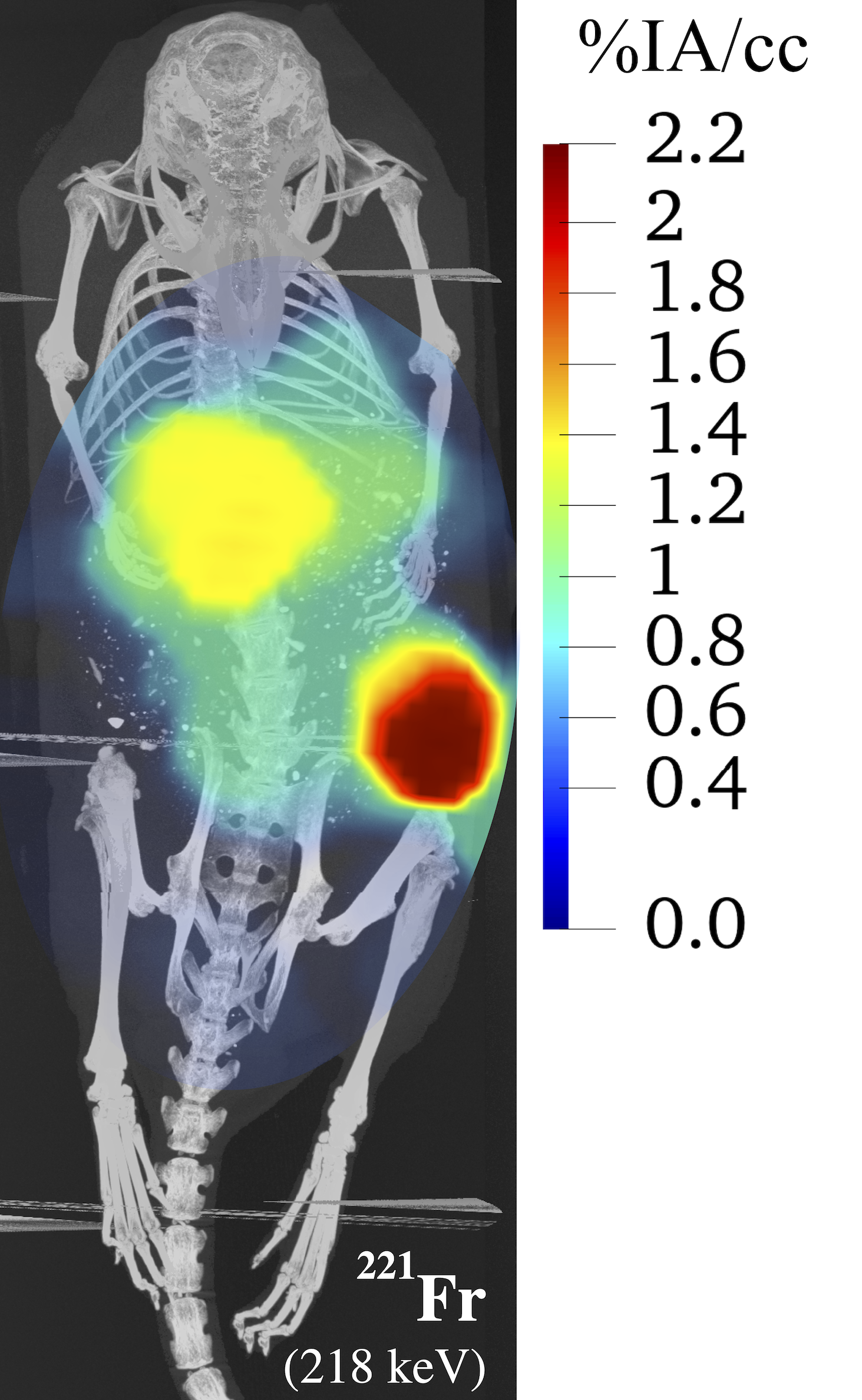}}
  \hspace{0.2cm}
  \subfloat[MIP CT + CI]{\includegraphics[width=0.2775\columnwidth]{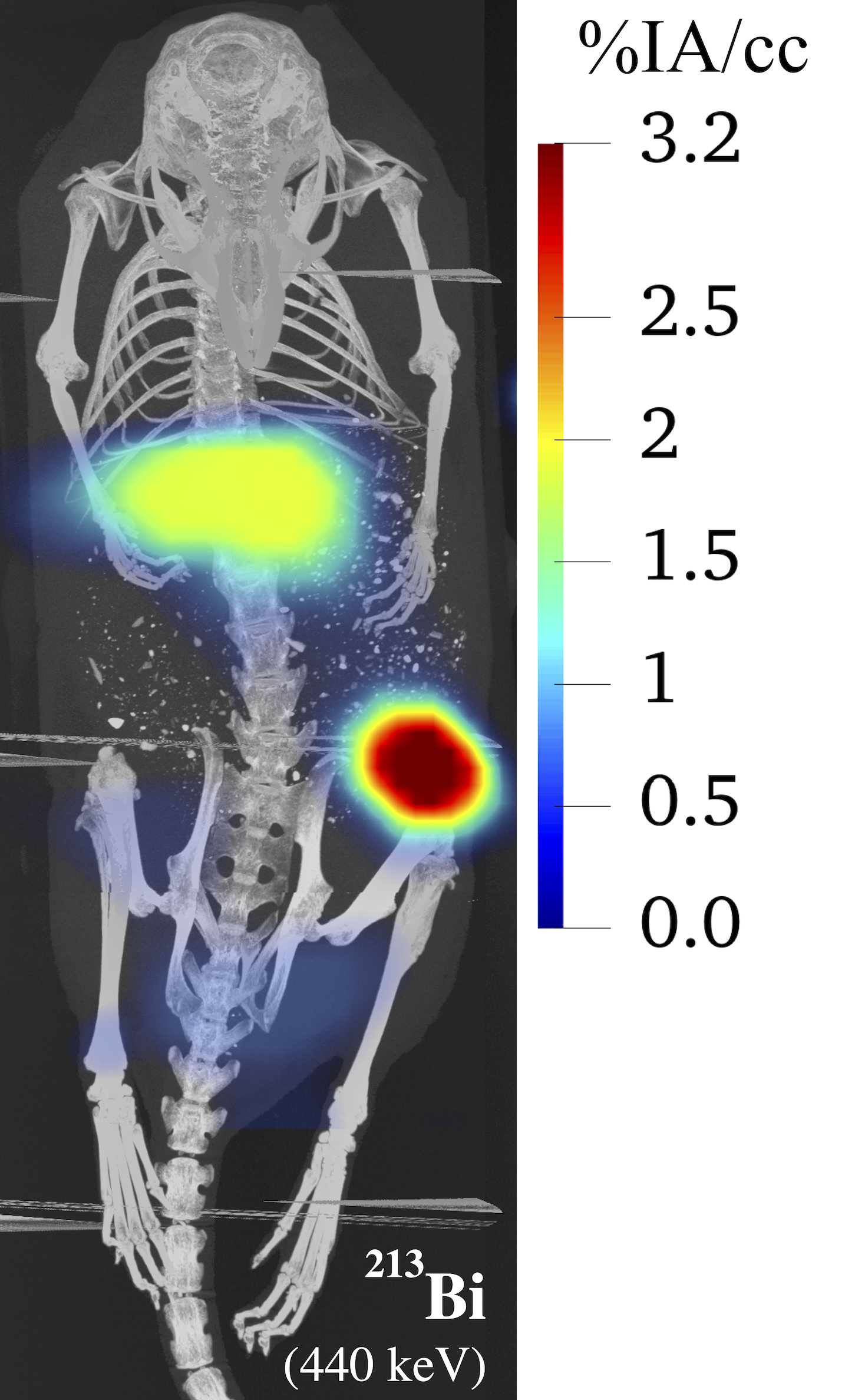}} \\
  \vspace{0.2cm}
  \rotatebox{90}{\quad\quad\footnotesize$^{225}$Ac-DOTA-YS5}
  \subfloat[2-D CT]{\includegraphics[width=0.1575\columnwidth]{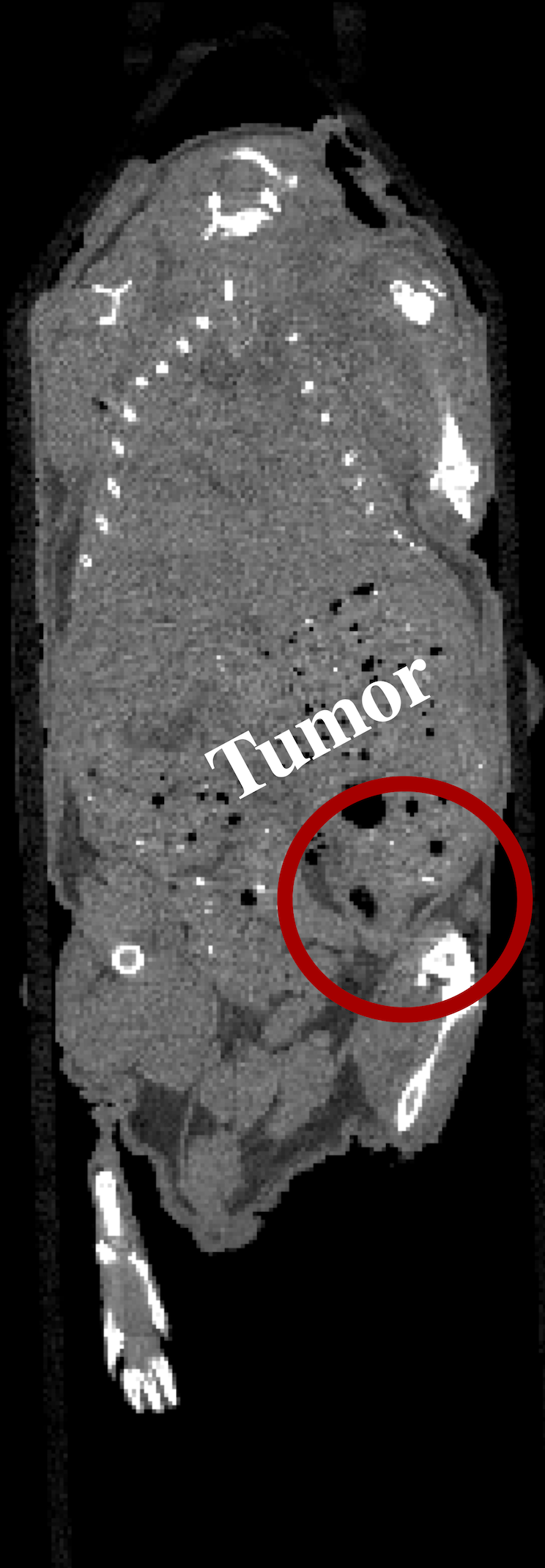}}
  \hspace{0.6cm}
  \subfloat[2-D CT + CA]{\includegraphics[width=0.279\columnwidth]{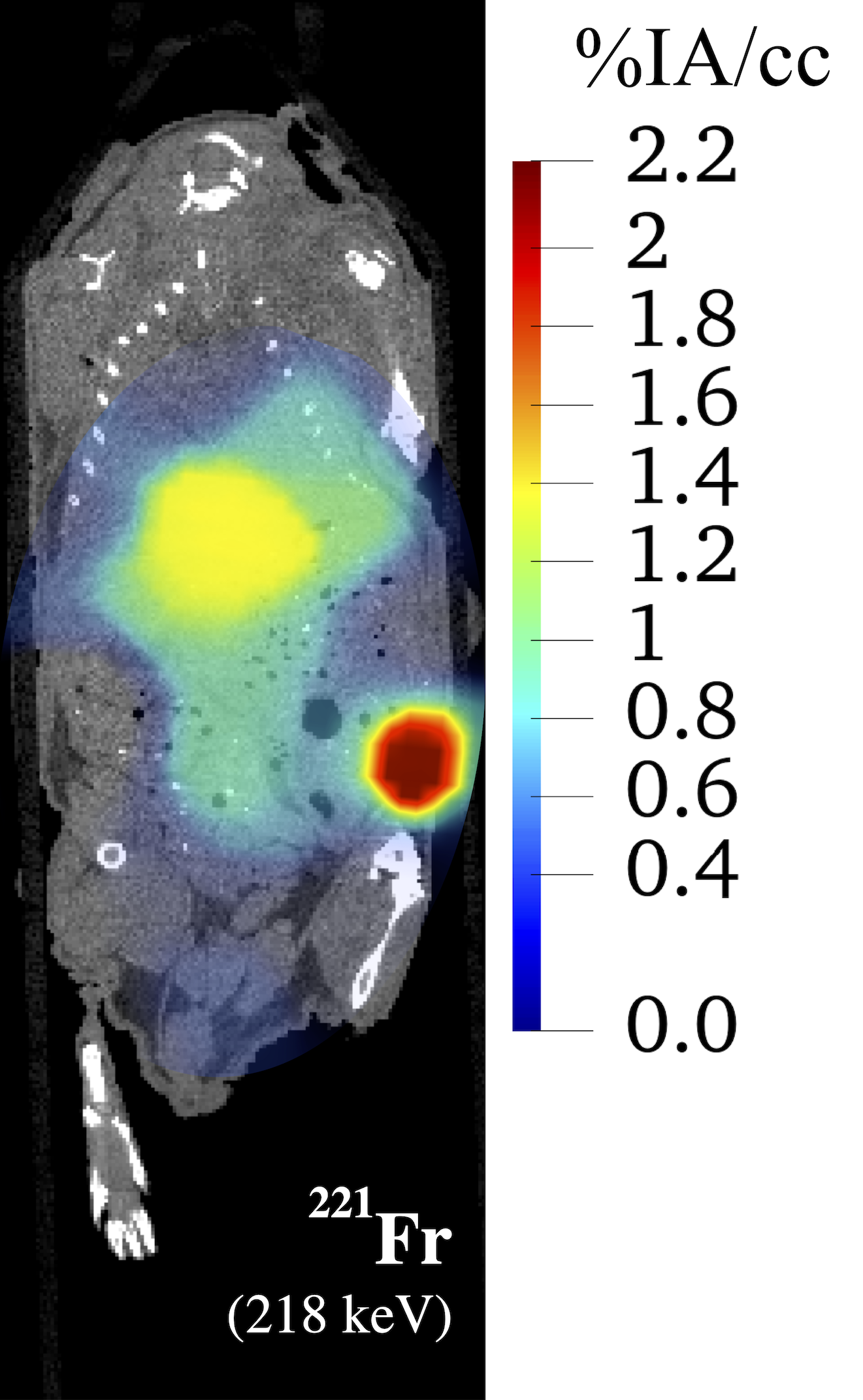}}
  \hspace{0.3cm}
  \subfloat[2-D CT + CI]{\includegraphics[width=0.2787\columnwidth]{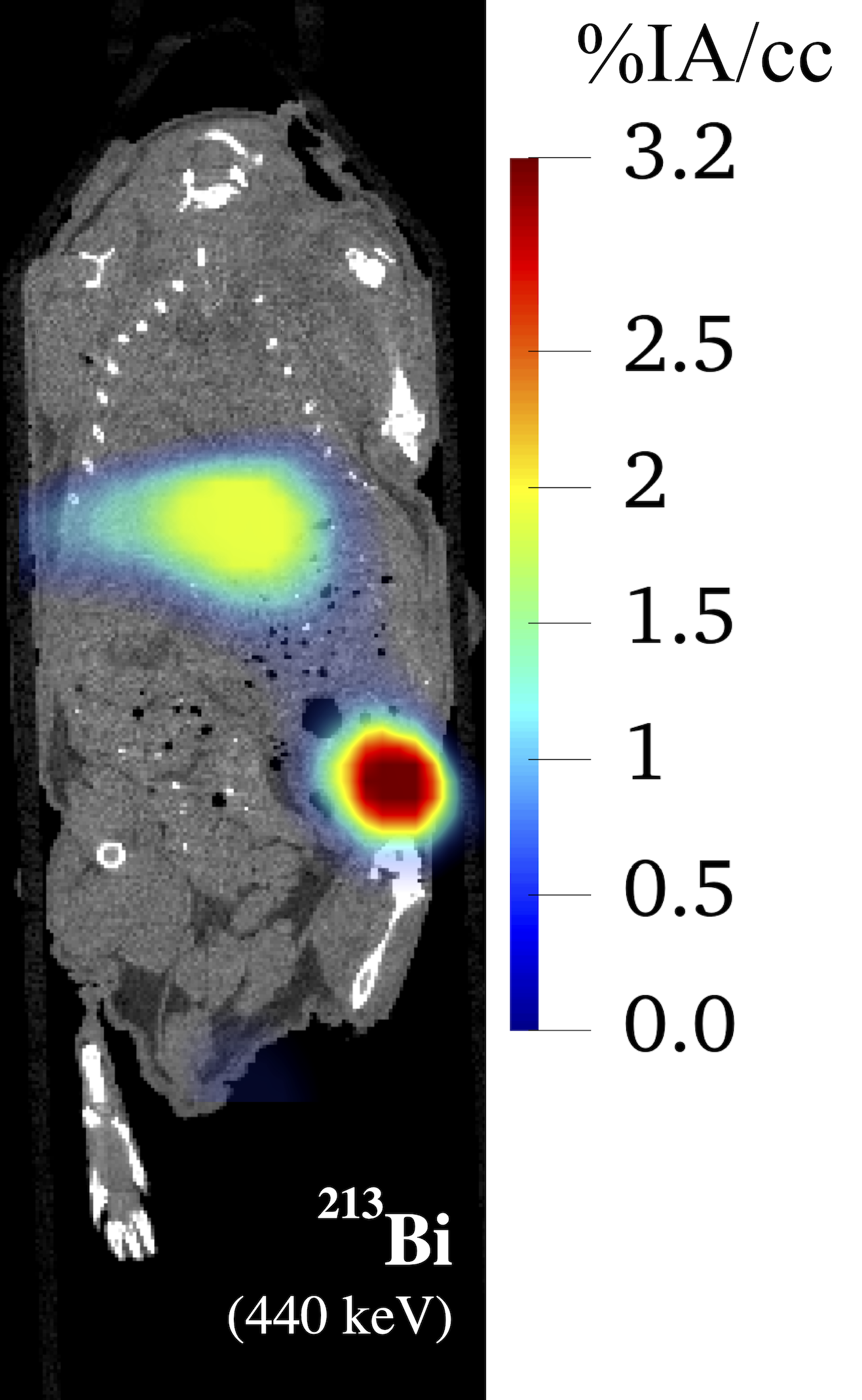}}
  \caption{(a) CT MIP of mouse B fused with the (b) coded aperture (CA) and (c) Compton (CI) MIPs. (d) CT coronal slice of mouse B fused with coronal slices of the (e) CA and (f) CI images. The CA image was generated from a total of $3.1 \times 10^4$ events at $218$~keV, and the CI image was generated from a total of $1.2 \times 10^4$ events at $440$~keV. The intensity scales are decay corrected to the day of injection.}
  \label{fig:8}
\end{figure}

Using ML-EM with TV, the two mice were reconstructed in 3-D from the coded aperture and Compton data at the $218$-keV emission line of $^{221}$Fr and $440$-keV emission line of $^{213}$Bi, respectively. Both the coded aperture and Compton images were generated from all eight projections after $250$~iterations. Figs. \ref{fig:7}b-c and \ref{fig:8}b-c show the maximum intensity projections (MIP) of the gamma-ray images of mice A and B, respectively. These images are fused with a CT MIP. Furthermore, Figs. \ref{fig:7}e-f and \ref{fig:8}e-f show coronal slices of the gamma-ray images of mice A and B, respectively. These images are fused with a coronal CT image. The quantification factor, determined from the phantom measurements in Section \ref{subsec:phantom}, was applied to each image so that the intensity scale is in units of percent injected activity per cubic centimeter (\%IA/cc). Additionally, the intensity scales are decay corrected to the day of injection.

Note the coded aperture and Compton images show no clear distinction between the central organs. This can be partially explained by the limited resolution of the imaging modalities; and thus going forward, improvements in resolution should be a priority. Furthermore, due to the limited and distinct resolution profiles of the coded aperture and Compton modalities, we cannot provide a concrete explanation for the discrepancies between the small-animal images of $^{221}$Fr and $^{213}$Bi. The variations are most likely imaging artifacts, but there could also be a physiological explanation.

\textit{Ex-vivo} biodistribution analyses of mice A and B were performed at $9$ and $11$~days post-injection, respectively, to validate the gamma-ray images. Major organs were harvested, weighed, and counted by a Hidex Automated Gamma Counter. Table \ref{tab:2} displays the activity estimates of $^{221}$Fr and $^{213}$Bi in the various organs as determined by the biodistribution analyses and coded aperture and Compton images. All activities are decay corrected to the day of injection.

\begin{table}[!tb]
  \begin{center}
  \caption{ Activity estimates in major organs of (a) mouse A and (b) mouse B as determined by the biodistribution (BioD) and coded aperture (CA) and Compton (CI) images. The biodistribution data are displayed as both activity in units of Becquerel (Bq) and percent injected activity per gram (\%IA/g). The image data are displayed as activity (Bq). The central organs encompass the liver, heart, and lungs. All reported values are decay corrected to the day of injection.} \label{tab:2}
  \subfloat[Mouse A injected with $20$~kBq of $^{225}$Ac-Macropa-PEG8(7)-YS5 \newline and sacrificed at $2$~days post-injection.]{
  \begin{tabulary}{\textwidth}{lrrrrrr}
  \toprule
  & & \multicolumn{2}{c}{BioD} & \multicolumn{1}{c}{CA} & \multicolumn{1}{c}{CI} \\
  \cmidrule(lr){3-4}
  \cmidrule(lr){5-5}
  \cmidrule(lr){6-6}
  Organ &
  Wt. &
  $^{221}$Fr &
  $^{213}$Bi &
  $^{221}$Fr &
  $^{213}$Bi \cr
  &
  [g] &
  [Bq, \%IA/g] &
  [Bq, \%IA/g] &
  [Bq] &
  [Bq] \cr
  \midrule
  Tumor & 0.300 & 1152, 19.2 & 1064, 17.8 & 1219 & 1035 \cr
  Liver & 0.597 & 786, 6.58 & 706, 5.91 & -- & -- \cr
  Heart & 0.252 & 246, 4.89 & 239, 4.74 & -- & -- \cr
  Lungs & 0.174 & 176, 5.06 & 195, 5.59 & -- & -- \cr
  Kidneys & 0.327 & 192, 2.93 & 210, 3.21 & -- & -- \cr
  Central & 1.02 & 1208, 5.90 & 1140, 5.57 & 1298 & 1220 \cr
  Organs \cr
  \bottomrule
\end{tabulary}
} \\
\subfloat[Mouse B injected with $20$~kBq of $^{225}$Ac-DOTA-YS5 and sacrificed at \newline $4$~days post-injection]{
  \begin{tabulary}{\textwidth}{lrrrrrr}
  \toprule
  & & \multicolumn{2}{c}{BioD} & \multicolumn{1}{c}{CA} & \multicolumn{1}{c}{CI} \\
  \cmidrule(lr){3-4}
  \cmidrule(lr){5-5}
  \cmidrule(lr){6-6}
  Organ &
  Wt. &
  $^{221}$Fr &
  $^{213}$Bi &
  $^{221}$Fr &
  $^{213}$Bi \cr
  &
  [g] &
  [Bq, \%IA/g] &
  [Bq, \%IA/g] &
  [Bq] &
  [Bq] \cr
  \midrule
  Tumor & 0.186 & 217, 5.84 & 198, 5.33 & 299 & 242 \cr
  Liver & 0.731 & 240, 1.64 & 238, 1.63 & -- & -- \cr
  Heart & 0.153 & 75, 2.46 & 71, 2.33 & -- & -- \cr
  Lungs & 0.199 & 122, 3.06 & 125, 3.14 & -- & --\cr
  Kidneys & 0.423 & 111, 1.32 & 94, 1.11 & -- & --\cr
  Central & 1.08 & 437, 2.02 & 434, 2.00 & 610 & 575 \cr
  Organs \cr
  \bottomrule
\end{tabulary}
}
\end{center}
\end{table}

\section{Discussion}
\label{sec:diss}
The ability to image the $^{225}$Ac daughters, namely $^{221}$Fr and $^{213}$Bi, in small animals would expedite the development of $^{225}$Ac-based radiopharmaceuticals, which have been demonstrated as a promising cancer treatment \cite{pros1}. This is the first time that $^{221}$Fr and $^{213}$Bi have been imaged via their gamma-ray emissions in mice using coded aperture and Compton methods. That said, a previous work by de Swart et al. \cite{swart} imaged $^{213}$Bi via its $440$-keV gamma-ray emission in a mouse using a multi-pinhole SPECT imager. The multi-pinhole system was designed to maximize resolution for energies up to $600$~keV at the expense of imaging sensitivity. Indeed, de Swart presents high-resolution images of $^{213}$Bi agents in mice. However, these results are based on an injected activity of $3$~MBq. This activity is $150$~times greater than that required of preclinical TAT studies. In contrast, we present images of mice with injected activities of $20$~kBq.

Other works have already proposed a Compton camera for the preclinical evaluation of the daughter $^{221}$Bi \cite{yoon, javier}; however, none have demonstrated the feasibility of this modality via experimentation. The most recent study by Caravaca et al. \cite{javier} simulates the response of a Compton imager based on CZT. The simulated results show the ability of the Compton camera to image $^{213}$Bi via its $440$-keV gamma-ray emission in a mouse phantom injected with $20$~kBq. Furthermore, Caravaca proposes a proximity imager to image $^{221}$Fr via its $218$-keV gamma-ray emission and presents simulated results. These results demonstrate the ability of the proximity modality to image $^{221}$Fr in a point-like phantom at a standoff distance of $1$~mm. In small-animal imaging, however, standoff distances are on the order of the extent of the animal body. At such distances, proximity imaging has not been demonstrated to have reasonable resolution.

Previous works have proposed and employed coded apertures for small-animal imaging \cite{accorsiSAMI}, but as far as we know, this work is the first to recommend a coded aperture for imaging $^{221}$Fr via its $218$-keV emission in small animals. Other analogous collimator-based systems have been employed to image $^{221}$Fr \cite{robertson, cai}. These studies show preliminary images of phantoms, but none have demonstrated sufficient sensitivity to image activities as low as $20$~kBq.

A study by Wang et al. \cite{wang2021} presents both a biodistribution analysis and PET image of a [$^{89}$Zr]DFO-YS5 agent in a mouse implanted with the same 22Rv1 prostate cancer cells as the mice presented here. The $^{89}$Zr agent is similar to the $^{225}$Ac-Macropa-PEG8(7)-YS5 and $^{225}$Ac-DOTA-YS5 agents injected in mice A and B, respectively. The results in Wang provide an estimate of the $^{225}$Ac biodistribution, which can serve as a foundation for assessing the daughter redistribution.

The biodistribution analysis of the [$^{89}$Zr]DFO-YS5 agent shows a tumor uptake of $14.5\pm3.2$~\%IA/g at $4$~days post-injection and less than $5$~\%IA/g in all other organs. In this work, the biodistribution of mouse A shows $19.2$~\%IA/g of $^{221}$Fr and $17.8$~\%IA/g of $^{213}$Bi in the tumor at $2$~days post-injection and less than $7$~\%IA/g of both isotopes in all other organs (Table \ref{tab:2}a). The biodistribution of mouse B shows significantly lower uptake. The tumor has $5.84$~\%IA/g of $^{221}$Fr and $5.33$~\%IA/g of $^{213}$Bi at $4$~days post-injection, and all other organs have less than $4$~\%IA/g of both isotopes (Table \ref{tab:2}b). The lower uptake could be related to a number of factors, including biological variations and a suboptimal tumor dissection. In regards to the latter, if the harvested tumor also included adjacent skin or muscle, then the uptake would appear less. This is a likely cause given that organs were found damaged upon dissection.

Similarly to the biodistribution analyses, the PET image of $^{89}$Zr (Fig. $3$a in Wang) and the gamma-ray images of $^{221}$Fr and $^{213}$Bi (Fig. \ref{fig:7}, Fig. \ref{fig:8}) show the highest uptake in the tumor and noticeable uptake in the central organs. The PET image of $^{89}$Zr has an intensity scale that ranges up to $20$~\%IA/cc, whereas the coded aperture and Compton images have intensity scales with maximums between $2$~\%IA/cc and $11$~\%IA/cc. The intensity variations between the PET image of $^{89}$Zr and the gamma-ray images of $^{221}$Fr and $^{213}$Bi might be indicative of the daughters redistributing, but for now, such claims cannot be made. The discrepancies might also be a result of biological variations associated with using both different mice and radioimmunotherapy agents. A more concrete explanation is the distinct resolution profiles of the imaging modalities. The coded aperture and Compton imagers have resolutions that are a factor of $5$ and $3$~times worse than a small-animal PET scanner, respectively, and thus the intensities of the coded aperture and Compton images spread over more image voxels. Going forward, the resolutions of the two modalities will need to match that of PET so that the daughter redistribution can be properly assessed.

The consistency between the biodistribution analyses and coded aperture and Compton images contributes to the credibility of this work. In the case of mouse A (Table \ref{tab:2}a), the biodistribution shows the highest uptake in the tumor with a total activity accumulation of about $1152$ Bq of $^{221}$Fr and $1064$ Bq of $^{213}$Bi. This is consistent with the coded aperture and Compton images, which show total accumulations of about $1219$ Bq of $^{221}$Fr and $1035$ Bq of $^{213}$Bi, respectively. Furthermore, the biodistribution of mouse A shows significant uptake in the liver, heart, and lungs. In combination, these three organs show total activity accumulations of about $1208$ Bq of $^{221}$Fr and $1140$ Bq of $^{213}$Bi. This is consistent with the coded aperture and Compton images, which show accumulations of about $1298$ Bq of $^{221}$Fr and $1220$ Bq of $^{213}$Bi, respectively, in the central region of the mouse.

In the case of mouse B (Table \ref{tab:2}b), the biodistribution shows the highest uptake in the tumor with a total activity accumulation of about $217$~Bq of $^{221}$Fr and $198$~Bq of $^{213}$Bi. This is consistent with the coded aperture and Compton images, which show total accumulations of about $299$~Bq of $^{221}$Fr and $242$~Bq of $^{213}$Bi, respectively. Furthermore, the biodistribution of mouse B shows significant uptake in the liver, heart, and lungs. In combination, these three organs show total activity accumulations of about $437$~Bq of $^{221}$Fr and $434$~Bq of $^{213}$Bi. The coded aperture and Compton images show accumulations of about $610$~Bq of $^{221}$Fr and $575$~Bq of $^{213}$Bi, respectively, in the central region. Note the biodistribution of mouse B shows significantly lower activities in the central organs compared to the images. This discrepancy can be partially explained by a suboptimal biodistribution analysis. Upon dissection of mouse B, several organs were found to be damaged and may not have been fully harvested.

While the quantification estimates from the biodistribution analyses and small-animal images are in close agreement, these estimates have no associated uncertainties. Going forward, a more sophisticated and comprehensive approach for quantifying the activities and estimating the associated uncertainties should be developed.

Another area of improvement is on image resolution. The current system could not discriminate the central organs, e.g. liver, heart, and lungs, given the $6.9$~mm and $4.1$~mm resolutions of the coded aperture and Compton modalities, respectively. Going forward, the image resolution can be improved significantly by reducing the strip pitch of the detectors and the size of the mask elements. For example, by using a $0.5$-mm strip pitch with mask elements to match, the coded aperture resolution would improve by about $75\%$ at the current magnification of $1.5$, and the Compton resolution would improve by about $40\%$ at a photon energy of $440$~keV and source-to-detector distance of $35$~mm. Additional improvements to the coded aperture resolution can be realized by increasing the magnification factor. If a magnification of $3$ were realized, then the coded aperture resolution would improve by an additional $50\%$. The Compton resolution can be further improved by decreasing the source-to-detector distance. If a distance of $20$~mm were realized, then the Compton resolution would improve by an additional $40\%$ \cite{frame2}.

Improvements to imaging sensitivity should also be a priority moving forward, so that the imaging time can be reduced and organs with relatively low uptake (e.g. kidneys) can be resolved. Currently, the coded aperture modality has a sensitivity of about $160$ cps/MBq ($0.016$\%) for $^{221}$Fr at a standoff distance of $145$~mm, and the Compton modality has a sensitivity of about $150$ cps/MBq ($0.015$\%) for $^{213}$Bi at a standoff distance of $35$~mm. These values can be improved by increasing the solid angle coverage of the detectors. For example, by simply reducing the standoff distance by half and adding four panels of each modality around the small animal, the sensitivity of both modalities would increase by a factor of $16$. Further sensitivity improvements can be realized by mitigating information loss in signal processing. This can be achieved by reducing the granularity of the detectors via a smaller strip pitch and employing more advanced signal processing algorithms.

\section{Conclusion}
\label{sec:concl}
The significant contribution of this work is the application of the coded aperture and Compton imaging concepts to the preclinical evaluation of the daughter redistribution of $^{225}$Ac-based radiopharmaceuticals. Up until now, there has been no effective means to study the daughter redistribution in small animals, which is an integral step towards clinical approval. This work presented coded aperture images of $^{221}$Fr and Compton images of $^{213}$Bi in both an $^{225}$Ac-filled phantom and tumor-bearing mice injected with $^{225}$Ac agents. Activity estimates were extracted from the small-animal images and validated by \textit{ex-vivo} biodistribution analyses. These results are the first demonstration of visualizing and quantifying the $^{225}$Ac daughters in small animals via coded aperture and Compton imaging.

\section*{Acknowledgements}
\label{sec:ack}
The authors would like to thank Ross Barnowski of the Dept. of Nuclear Engineering at UC Berkeley for his contribution in the development of the Dual-Modality Imager, Rebecca Abergel and Lee Bernstein of the Dept. of Nuclear Engineering at UC Berkeley for their valuable insight on TAT, Bin Liu of the Dept. of Anesthesia at UCSF for developing the YS5 antibody, Ian Baldridge of EH\&S at UC Berkeley for delivering the mice, Ryan Tang and Youngho Seo of the Dept. of Radiology and Biomedical Imaging at UCSF for providing the CT data, and finally the US Department of Energy Isotope Program for providing the $^{225}$Ac.

\end{document}